\documentclass[journal]{IEEEtran}

\usepackage{epsf,psfrag,amssymb,amsfonts,color,cite,fancybox}

\usepackage[mathscr]{eucal}
\usepackage[dvips]{graphicx}
\usepackage[dvipsnames]{xcolor}
\usepackage{caption}
\usepackage{ifpdf}
\usepackage{longtable}	
\usepackage{multicol}
\usepackage{float}
\usepackage{epstopdf}
\usepackage{bm}
\usepackage{multirow}
\usepackage[scientific-notation=true]{siunitx}
\usepackage{url}
\usepackage{cancel}
\usepackage{ulem}
\usepackage[dvipsnames]{xcolor}
\usepackage{subfigure}

\usepackage{algorithm}
\usepackage{algpseudocode}
\usepackage{makecell}

\ifCLASSINFOpdf

\else

\fi

\usepackage[cmex10]{amsmath}
\begin{document}
\bstctlcite{IEEEexample:BSTcontrol}
\title{Targeted False Data Injection Attacks Against AC State Estimation Without Network Parameters}

\author{Mingqiu Du,~\IEEEmembership{Graduate Student Member,~IEEE}, Georgia Pierrou, ~\IEEEmembership{Student Member,~IEEE}, Xiaozhe Wang,~\IEEEmembership{Senior Member,~IEEE}, Marthe Kassouf

\thanks{This work is supported by Natural Sciences and Engineering Research Council (NSERC) Discovery Grant RGPIN-2016-04570 and Mitacs Accelerate IT22895.}
\thanks{Mingqiu Du, {Georgia Pierrou}, Xiaozhe Wang are with the Department of Electrical and Computer Engineering, McGill University, Montr\'{e}al, QC H3A 0G4, Canada. (email: {mingqiu.du@mail.mcgill.ca, georgia.pierrou@mail.mcgill.ca}, xiaozhe.wang2@mcgill.ca)}% <-this % stops a space
\thanks{ Marthe Kassouf is with Hydro-Quebec Research Institute (IREQ) (email: kassouf.marthe@ireq.ca)}}

\maketitle

%%%************************************************************************************
%%%*** Abstract and Keywords                                                        ***
%%%************************************************************************************

%
\begin{abstract}
State estimation is a data processing algorithm for converting redundant meter measurements and other information into an estimate of the state of a power system. Relying heavily on meter measurements, state estimation has proven to be vulnerable to cyber attacks. \color{black}In this paper, a {novel} targeted false data injection attack (FDIA) model against AC state estimation is proposed. Leveraging on the intrinsic load dynamics in ambient conditions and important properties of the Ornstein-Uhlenbeck process, we, from the viewpoint of intruders, design an algorithm to extract power network parameters purely from PMU data, which are further used to construct the FDIA vector. \color{black} Requiring no network parameters and relying only on limited phasor measurement unit (PMU) data, the proposed FDIA model can target specific states and launch large deviation attacks. Sufficient conditions for the proposed FDIA model are also developed. Various attack vectors and attacking regions are studied in the IEEE 39-bus system, showing that the proposed FDIA method can successfully bypass the bad data detection and launch targeted large deviation attacks with very high probabilities. 
\end{abstract}


\IEEEpeerreviewmaketitle

%%%************************************************************************************
%%%*** Introduction                                                                 ***
%%%************************************************************************************

\section{Introduction}\label{introduction}
\IEEEPARstart {E}{nsuring} the security of electric power systems is of paramount importance to economic prosperity, normal operation of a society, and even national security. The state estimator, being the core component in the supervisory control and data acquisition (SCADA) system, constantly monitors the operating state of a power system for achieving a safe and reliable operation. However, the state estimator is subject to cyber-attacks due to the integration of information technology and the vulnerability of communication networks.

In fact, it has been shown in %Liu \emph{et al}.
\cite{Liu2009} that the intruder can compromise the state estimator by injecting pre-designed false data into meters without being detected by the bad data detection (BDD), provided that the full topology and line parameters of a power grid are known to the intruder. Inspired by this important work, many subsequent efforts have been made to investigate false data injection attack (FDIA) on state estimation of power grids. 
At the early stage, DC state estimation model is typically exploited due to its simplicity and linearity (e.g. \cite{Liu2009}, \cite{Rahman2012}). 
However, it has been shown that the FDIA designed based on DC model can be easily detected if AC model is implemented, which is most likely the case in practical applications \cite{Rahman2013}. 

Designing FDIA against AC state estimation seems challenging due to the nonlinearity and non-existence of analytical solution of AC state estimation problem. The authors of \cite{Hug2012} developed an analytical technique for performing vulnerability analysis of AC state estimation under FDIA. Analytical bounds on the uncertain information about a successful FDIA against AC state estimation have been shown in \cite{Zhao2018}. Another pioneer work about FDIA on AC state estimation \cite{Rahman2013} studied the possibility of FDIA against AC state estimation with perfect and imperfect knowledge of states.   Recently, FDIA against AC state estimation in distribution networks has been studied in \cite{Deng2019a}. The radial structure of the distribution network was leveraged to reduce the number of states known to the intruder. To minimize the number of compromised measurements, optimization problems were formulated in \cite{Jin2017, Nayak2020,Sawas2018, Liu2021}  to incorporate constraints of limited resources and stealth operation. Semidefinite programming (SDP) relaxation and a sparsity penalty were exploited to recover a near-optimal solution in \cite{Jin2017}. A reduced row-echelon (RRE) form-based greedy method was utilized in \cite{Nayak2020} to compute the minimum number of measurements. The genetic algorithm and a neural network were applied in \cite{Sawas2018} to construct the least-effort attack vector. The manipulation of network parameters in addition to system states was proposed in \cite{Liu2021} to reduce the total number of compromised measurements. In addition, the impact analysis about FDIAs against AC state estimation was carried out in {electricity} market \cite{Xie2010} and load redistribution  \cite{Yuan2011a}, both causing huge financial losses. 

Despite significant advancements in the study of FDIA against AC state estimation, all the aforementioned works assume that the intruder has perfect knowledge (e.g., line admittance, topology) of the whole transmission network. To relax this assumption, the concept of attacking region (i.e., partial power network where attacks can be launched) originally proposed in DC model\cite{Rahman2012} has been exploited in \cite{Liu2017, Deng2019}, in which only partial line admittance and topology are needed.  Particularly, the FDIA model in \cite{Liu2017} requires only the line admittance inside the attacking region. The FDIA model developed in \cite{Deng2019} needs only the susceptance of all lines connected to the target bus or the superbus (a set of interconnected {buses}). 

However, the required partial transmission network information may still be unobtainable as the system model is well protected by utilities. Until recently, a network-independent FDIA is proposed in \cite{Chin2018} using principal component analysis (PCA) and geometric approach, which requires no transmission network information. However, the proposed FDIA model cannot target specific measurements, while large deviation attacks may not be launched due to the enforced DC approximations.

In this paper, we propose a novel targeted FDIA model against AC state estimation. Requiring no line parameters of a power network and leveraging purely on PMU measurements, the proposed FDIA model can target specific states and launch large deviation attacks. As measurement data of widely deployed PMUs is relatively easier to acquire compared to the power network information critically protected \cite{Beasley2015}, we, from the perspective of the intruder, utilize PMU measurements inside the attacking region to design a stealthy FDIA against AC state estimation.  The aforementioned PMU measurements may be intercepted in the communication network \cite{Beasley2015} or infiltrated by leveraging software supply chain vulnerabilities such as for the SolarWinds attack \cite{article}. Once the PMU data is collected, we leverage the intrinsic load dynamics inside the attacking region and draw upon the regression theorem of the Ornstein-Uhlenbeck process and weighted least square estimations to estimate the parameters of the network needed to construct the attack vector. It should be noted that no additional measurements other than the aforementioned limited PMU data are needed. 

To our best knowledge, the proposed attack model requires the least amount of measurements for designing the FDIA against AC state estimation that can target specific states and launch large deviation attacks without any line parameters of a network. The advantages of the proposed FDIA model against AC state estimation over existing works and the contributions of the paper are summarized below.

\begin{itemize}
\item Compared to the FDIA models designed in  \cite{Rahman2013,Hug2012,Zhao2018,Deng2019,Deng2019a,Jin2017,Liu2021,Sawas2018,Liu2017,Nayak2020},  the proposed FDIA model requires no line parameters. 

\item Compared to the PCA-based network-independent FDIA model designed in \cite{Chin2018}, the proposed FDIA model can target specific states and launch large deviation attacks, leading to more specific or more severe consequences. 
\item The proposed FDIA model, assuming no DC approximations (e.g., $\cos ({\delta _i} - {\delta _j}) \approx 1; \sin ({\delta _i} - {\delta _j}) \approx {\delta _i} - {\delta _j}$; ${V_i} \approx {V_j} \approx 1$), can bypass the BDD with a very high probability even under large deviation attacks as shown in the numerical simulations. 
\item 
Sufficient conditions for the proposed FDIA model to be perfect, i.e., the residual remains the same before and after the attack, are given, providing a theoretical basis for the proposed FDIA model.  
\end{itemize}

The proposed FDIA model against state estimation indicates that it may be even easier than we thought to launch an FDIA, as only limited PMU measurements are needed without any line parameters of a power network. Designing countermeasures against this FDIA model is significant and imperative in the near future.

\section{A Brief Review of  FDIA against AC State Estimation }\label{Background for the FDIA}

\noindent$\bullet$ \textbf{State Estimation:} In power system operation, SCADA system is responsible for transmitting measurements, status information, and circuit-breaker signals to and from remote terminal units (RTUs). The relationship between measurements $\bm{z}$ and state variables $\bm{x}$ can be formulated using the nonlinear equation $\bm{h}$ as follows:  
\begin{equation}
\label{relationship between z and x}
\bm{z} = \bm{h}(\bm{x}) + \bm{e}
\end{equation}
where the state vector $\bm{x}$ includes bus voltage angles and magnitudes to be estimated; $\bm{z}$ consists of power injections and bidirectional power flows for each line  acquired from RTUs ; $\bm{e}$ is the vector of measurement error, which is typically assumed to be an independent Gaussian random vector. To get the estimated state vector $\hat{\bm{x}}$, the least squares estimation problem is formulated as 
$\bm{\hat x} = \arg \min_{\bm{x}} \left\| {\bm{e}} \right\|_2 = \arg \min_{\bm{x}} \left\| {\bm{z} - \bm{h}(\bm{x})} \right\|_2$ \cite{Monticelli1999}.  Although PMU data can also be used for state estimation, it is assumed that only RTU measurements of line flow and power injections are exploited for state estimation similar to  \cite{Rahman2012,Rahman2013,Liu2015}.

\noindent $\bullet$ \textbf{Bad Data Detection:} 
When collecting the measurements in $\bm{z}$, bad data may appear due to meter failure, communication interruption,  malicious attack, etc. To detect bad data, the residual-based BDD is typically used \cite{Monticelli1999}. 
The residual between the measurements and the calculation using the estimated state vector is defined as follows:

\begin{equation}
\label{residual_test_infinite}
r = {\left\| {\bm{z} -{\bm{h}}(\bm{\hat x})} \right\|_\infty } = \mathop {\max }\limits_i \left| {{\bm{r}_i}} \right|
\end{equation}
\normalsize

If $r$ is less than a predetermined threshold value $\gamma $, the measurements are assumed to be normal; otherwise, measurements are considered to contain bad data by the BDD and an alarm will show up in the control room. 

\noindent $\bullet$ \textbf{False Data Injection Attack:}
We define ${\bm{z}_{bad}}$ as the measurement vector that contains malicious data. Then, ${{\bm{z}_{bad}}}$ can be represented as ${\bm{z}_{bad}} = \bm{z} + \bm{a}$, where $\bm{z}$ represents the original measurement vector and $\bm{a}$ represents the attack vector intentionally added to the original measurement $\bm{z}$.
Let ${{{\bm{\hat x}}_{bad}}}=\hat{\bm{x}}+\bm{c}$ denote the false state vector that the intruder intends to inject and %estimation of $\bm{x}$ after the attack and 
let $\bar{\bm{x}}$ be the point that the numerical method applied in the AC state estimation converges to after the attack. %using ${\bm{z}_{\bm{bad}}}$. 
It has been shown in \cite{Rahman2013} that if the attack vector is well designed such that $\bm{a}=\bm{h(\hat{\bm{x}}}+\hat{\bm{c}})-\bm{h(\hat{x})}$, then the residual after the attack remains the same:
\begin{eqnarray}
r_{bad}&=&\|\bm{z}+\bm{a}-\bm{h}(\bar{\bm{x}})\|_{\infty}\nonumber\\
&=&\|\bm{z}+\bm{h}(\bm{\hat{x}}+\bm{c})-\bm{h(\hat{x})}-\bm{h(\bar{x})}\|_{\infty}\nonumber\\
&=&\|\bm{z}-\bm{h(\hat{x})}+\bm{h}(\bm{\hat{x}}+\bm{c})-\bm{h(\bar{x})}\|_{\infty}\nonumber\\
&=&\|\bm{z}-\bm{h(\hat{x})}\|_{\infty}=r\label{eq:residual}
\end{eqnarray}
under the assumption that  $\bm{h(\bar{x})}=\bm{h}(\bm{\hat{x}}+\bm{c})$ is satisfied. %at each iteration of the numerical method applied. %in the AC state estimation. 
The authors of \cite{Rahman2013} show that it is true in most practical scenarios.

It is evident from the above analysis that in order to build the attack vector $\bm{a}$, %given the deviation vector $\bm{c}$ to be injected, 
the attacker needs to have perfect knowledge of the system states $\bm{x}$ and $\bm{h}$, i.e., the complete information of the power network topology and line parameters, which can be hardly acquired in practice. To relax the assumption of knowing the complete information of transmission network, %In particular to the AC model, 
the authors of \cite{Liu2017}, \cite{Wang2018a} utilized the concept of attacking regions and proposed FDIA models on AC state estimation that need only the network information \textit{inside} the attacking region.
Nevertheless, the required partial network information %inside the attacking region 
may still be unobtainable. In the next section, we will develop a novel FDIA against AC state estimation that can target specific states, requiring no line parameters, regardless of inside or outside the attacking region.

\section{Construction of the FDIA Model}\label{construction for FDIA}

To launch successful FDIA attacks, some common assumptions are typically made (e.g., \cite{Yuan2011a,Liu2014}) that we will also follow in this paper: 
\begin{itemize}
\item[C1.]\label{C1} Generator output measurements are not compromised because direct communications between power generators and the control center may exist, making attacks easily detected \cite{Ganugula2001}.

\item[C2.] \label{C2}Line flow and bus injection measurements at load buses can be compromised because load meters are widely distributed and load powers are constantly varying. 
\end{itemize}

\subsection{Defining the Attacking Region}\label{section_attackingregion}

\begin{figure}[htbp]
\includegraphics[width=0.45\textwidth,keepaspectratio=true,angle=0]{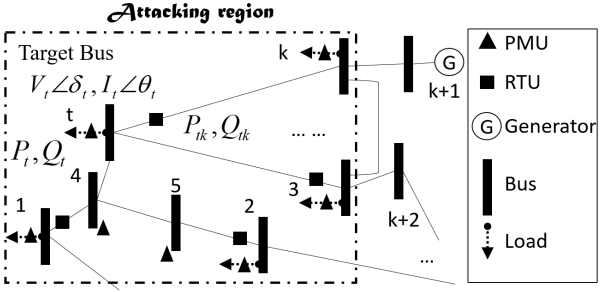}
\caption{Definition of the attacking region}
\label{attacking_region}
\end{figure}

Let $\Omega  = \{ 1,2,3,...,\}$ be the set of all power system buses and $L \subseteq \{ (p,q)|(p,q) \in \Omega  \times \Omega\mbox{, }p \ne q\} $ be the set of all transmission lines.
Inspired by the utilization of attacking region, assuming the intruder intends to attack the states of a particular bus (e.g., the target bus $t$ in Fig. 1), the attacking region can be defined as an undirected graph $S_A=(\Omega_A,L_A)$, where $\Omega_A$ is the set of power buses inside the attacking region and $L_A$ is the set of transmission lines inside the attacking region following the steps below. \\
a). Let ${\Omega _A} = \{ t\} $, ${L_A} = \phi $, ${\Omega_B} = \Omega_A \setminus \{t\}$. \\
b). Find all $i$ such that $(t,i)\in L$. 
%If $\exists i \in \Omega_A$ such that $(t,i)\in L$, find all $i$ %such that $(t,i)\in L$, 
Let ${\Omega _A} = {\Omega _A} \cup \{ i\} ,{L_A} = {L_A} \cup \{ (t,i), (i,t)\}$, ${\Omega_B} = \Omega_B \cup \{i\}$. \\
c). For all $j\in \Omega_B$, if $P_j\neq 0$, i.e., $j$ is a non zero-injection bus, stop; otherwise, for each $j$ that {$P_j=0$}, let $\Omega_B=\Omega_B\setminus\{j\}$, $t\leftarrow j$, and go back to b). 

Note that $\Omega_B$ corresponds to the set of boundary buses of the attacking region. Following the algorithm above, whenever a zero-injection bus is encountered and included in the boundary bus set, the attacking region expands to include its adjacent buses until a non zero-injection bus is included in the attacking region and serves as the boundary bus. The reason for doing this is that the total power flow of the zero-injection bus must be zero \cite{Hug2012,Zhao2018}. If one line flow of the non-zero injection bus is changed, then another line flow of the bus and thus the power injection measurement of its adjacent bus must be changed accordingly.

In addition, some assumptions inside the attacking region are as follows. 
\begin{itemize}

\item[A1.]\label{A1} Similar to the assumption adopted in previous works (e.g., \cite{Rahman2013,Liu2009,Deng2019}),  
the intruder can obtain the voltage and current phasors inside the attacking region through compromising PMUs. {However, other  RTU measurements (e.g., line flows and power injections) are not needed for constructing the attack vector.}

\item [A2.]\label{A2} Similar to 
\cite{Rahman2013,Yuan2011a,Liu2014}, the intruder can manipulate active power and reactive power injections of buses inside the attacking region (e.g., $P_t$, $Q_t$ in Fig. \ref{attacking_region}), the bidirectional  active line flows and reactive line flows (e.g., $P_{tk}$, $Q_{tk}$ and $P_{kt}$, $Q_{kt}$ in Fig. \ref{attacking_region}), and the PMU measurements {of voltage and current phasors} (e.g., $V_t\angle \delta_t$ and $I_t\angle \theta_t$  in Fig. \ref{attacking_region}) inside the attacking region. 

\end{itemize}

It is worth noting that for all lines in ${L_A}$, no parameters are required. Besides, the connecting status/topology {of any line that} does not belong to ${L_A}$ (e.g., the line between bus $k$ and bus 3) is not needed. Compared to the previous works \cite{Rahman2013,Liu2017}, pre-knowledge of detailed topology and line parameters of the attacking region and the boundary lines is significantly relaxed. 

In addition, it should be noted that because of C1 and A2, generator buses cannot be included in an attacking region. That being said, buses directly connected to the generator buses are not assumed to be attacked.

\subsection{Designing the Attack Vector}

As shown in (\ref{eq:residual}), the attack vector needs to satisfy $\bm{a} = \bm{h}(\bm{\hat x} + \bm{c}) - \bm{h}(\bm{\hat x})$ such that the residual after the attack remains the same as that before the attack. More specifically, denoting the set of buses inside the attacking region ${\Omega _A} = \{ 1,2,...,t,...,k\}$, if the intruder wants the control room to believe that the voltage of target node is ${\tilde V_t}\angle {\tilde \delta _t}$ rather than the true value ${V_t}\angle {\delta _t}$ in Fig. \ref{attacking_region}, we need to first determine the voltage phasors of the zero-injection buses. For illustration purpose{s}, consider the example in Fig. \ref{attacking_region}, the voltage phasors of the zero-injection buses, bus 5 and bus 4, need to be determined first by using the condition that the power injections are zero: 
\small

{\begin{align}
S_{5} = \tilde{V}_5\angle \tilde{\delta}_5(\frac{\tilde{V}_5\angle \tilde{\delta}_5-{V}_2\angle {\delta}_2}{R_{52}+jX_{52}})^{\star}+
\tilde{V}_5\angle \tilde{\delta}_5(\frac{\tilde{V}_5\angle \tilde{\delta}_5-\tilde{V}_4\angle \tilde{\delta}_4}{R_{54}+jX_{54}})^{\star}=0\nonumber\\
S_{4} = \tilde{V}_4\angle \tilde{\delta}_4(\frac{\tilde{V}_4\angle \tilde{\delta}_4-\tilde{V}_5\angle \tilde{\delta}_5}{R_{45}+jX_{45}})^{\star}+\tilde{V}_4\angle \tilde{\delta}_4
(\frac{\tilde{V}_4\angle \tilde{\delta}_4-{V}_1\angle{\delta}_1}{R_{41}+jX_{41}})^{\star}\nonumber\\
+\tilde{V}_4\angle \tilde{\delta}_4(\frac{\tilde{V}_4\angle \tilde{\delta}_4-\tilde{V}_t\angle\tilde{\delta}_t}{R_{4t}+jX_{4t}})^{\star}=0
\label{eq:S5S4}
\end{align}
}\normalsize
where $S$ denotes the complex power; {$R_{ij}$} and {$X_{ij}$} are resistance and reactance, respectively; $^\star$ denotes the conjugate operation. The states of the involved non zero-injection buses, i.e., bus 1, bus 2, are known from PMUs and {remain} unchanged after the attack. In fact, $\forall i \in {\Omega _B}$, ${{\tilde V}_i}\angle {{\tilde \delta }_i}$ after the attack has to remain the same as ${V_i}\angle {\delta _i}$ since the intruder cannot manipulate measurements outside the attacking region while power balance has to be satisfied. It can be seen that if the line parameters $R_{ij}$ and $X_{ij}$ are known, $\tilde{V}_4\angle \tilde{\delta}_4$ and $\tilde{V}_5\angle \tilde{\delta}_5$ can be solved from (\ref{eq:S5S4}). Next, the line flows inside the attacking region to be manipulated can be determined by:
\small
\begin{eqnarray}
\label{fake_P_ij}
{{\tilde P}_{ij}} = {{\tilde V}_i}{G_{ij}} - {{\tilde V}_i}{{\tilde V}_j}({G_{ij}}\cos {{\tilde \delta }_{ij}} + {B_{ij}}\sin {{\tilde \delta }_{ij}})\\
\label{fake_Q_ij}
{{\tilde Q}_{ij}} =  - {{\tilde V}_i}{B_{ij}} - {{\tilde V}_i}{{\tilde V}_j}({G_{ij}}\sin {{\tilde \delta }_{ij}} - {B_{ij}}\cos {{\tilde \delta }_{ij}})
\end{eqnarray}
\normalsize
$\forall (i,j)  \in {L_A}$. The power injections of non zero-injection buses to be manipulated can be determined by:
\small
\begin{eqnarray}
\label{boundary_P}
{{\tilde P}_i} = {P_i} + \sum\nolimits_{\{ i,j\}  \in {L_A}} {({{\tilde P}_{ij}} - {P_{ij}})}\\
\label{boundary_Q}
{{\tilde Q}_i} = {Q_i} + \sum\nolimits_{\{ i,j\}  \in {L_A}} {({{\tilde Q}_{ij}} - {Q_{ij}})}
\end{eqnarray}
\normalsize

Lastly, the current phasors of PMUs to be manipulated can be determined from: 
\begin{equation}
\label{fake_I_i}
{{\tilde I}_i}\angle {{\tilde \theta }_i} = (\frac{{\tilde P}_{i} + j{{\tilde Q}_{i}}}{{{\tilde V}_i}\angle  {{\tilde \delta }_i}})^{\star}
\end{equation}
$\forall i \in {\Omega _A}$. In other words, if $\bm{c}=[\tilde {V}_t-V_t, \tilde {\delta}_t-\delta_t]^T$, the attack vector $\bm{a}$ is designed as: 
\begin{equation}
\begin{array}{l}
\bm{a} = [{{\tilde P}_{ij}} - {P_{ij}},{{\tilde Q}_{ij}} - {Q_{ij}},{{\tilde P}_k} - {P_k},{{\tilde Q}_k} - {Q_k},\\
{{\tilde V}_m}\angle {{\tilde \delta }_m} - {V_m}\angle {\delta _m},{{\tilde I}_n}\angle {{\tilde \theta }_n} - {I_n}\angle {\theta _n}{]^T}
\end{array}
\label{eq:attack vector}
\end{equation}
$\forall (i,j) \in {L_A}, \forall k \in \Omega_C, \forall m \in {\Omega _A}\setminus \Omega_B, \forall n \in {\Omega_A}$, where $\Omega_C$ is the set containing all non zero-injection buses inside the attacking region. The attack vector designed in this way can successfully bypass the BDD according to (\ref{eq:residual}). Also, the power injections of the zero-injection buses inside the attacking region are ensured to be zero.

It is evident from (\ref{eq:S5S4})-(\ref{fake_Q_ij}) that in order to have the FDIA stealthily bypass the BDD, the intruder needs to have perfect knowledge of $G_{ij}$ and $B_{ij}$, $\forall (i,j)  \in {L_A}$,  %i.e., the line parameters inside the attacking region,
 as assumed in previous papers. However, such parameters inside the attacking region may still be hard to obtain as the system model is critically protected. {To relax this assumption, we leverage on the inherent load dynamics inside the attacking region and the regression theorem of the Ornstein-Uhlenbeck process to extract the required information of the physical system as discussed in the next section. }

\subsection{The Dynamic Model Inside the Attacking Region}

Inside an attacking region as shown in Fig. \ref{attacking_region}, the power system can be represented as a set of Differential-Algebraic equations as follows\cite{Kundur1993}:

\small
\begin{eqnarray}
\bm{\dot x} &=& \bm{f}(\bm{x},\bm{z})
\label{differentiable-algebraic eqn1}\\
\bm{z} &=& \bm{h}(\bm{x})
\label{differentiable-algebraic eqn2}
\end{eqnarray}
\normalsize
where (\ref{differentiable-algebraic eqn1}) describes %{generator dynamics}, 
load dynamics and (\ref{differentiable-algebraic eqn2}) describes the power flow relationship. In previous FDIA research on state estimation, only (\ref{differentiable-algebraic eqn2}) is considered as discussed in Section \ref{Background for the FDIA}, whereas the dynamics described by (\ref{differentiable-algebraic eqn1}) are neglected by assuming the system is operating in steady state. In the following, we will show that by exploring the load dynamics, essential and critical information about physical systems can be acquired.

In particular, for any bus $i$ inside the attacking region $\Omega_A$, the load dynamics can be represented in their detailed form:  

\small
\begin{eqnarray}
{\dot \delta _i} = \frac{1}{{{\tau _{{p_i}}}}}(P_i^s - {P_i})\label{ddelta_dP}\\
{\dot V_i} = \frac{1}{{{\tau _{{q_i}}}}}(Q_i^s - {Q_i})\label{dV_dQ}
\end{eqnarray}

\begin{eqnarray}
{P_i} = {V_i}\sum\limits_{j \in {\Omega _i}} {({V_j}{G_{ij}}\cos {\delta _{ij}} + {V_j}{B_{ij}}\sin {\delta _{ij}})} 
\label{bus active power injection}\\
{Q_i} = {V_i}\sum\limits_{j \in {\Omega _i}} {({V_j}{G_{ij}}\sin {\delta _{ij}} - {V_j}{B_{ij}}\cos {\delta _{ij}})} 
\label{bus reactive power injection}
\end{eqnarray}
\normalsize

\noindent where %\color{red} the space needs to be shrinked in some waydone\\
\begin{tabular}{cl}
  ${\Omega _i}$ & the set of buses that are connected to bus $i$\\
  $\delta_{i}$ & voltage angle of bus $i$\\
  $V_i$ & voltage magnitude of bus $i$\\
  ${P_{i}}$ & active power injection of bus $i$\\
  ${Q_{i}}$ & reactive power injection of bus $i$ \\
  ${P_i^s}$ & static active power injection of bus $i$\\
  ${Q_i^s}$ & static reactive power injection of bus $i$ \\
  ${P_{ij}}$ & active power flow from bus $i$ to $j$ \\
  ${Q_{ij}}$ & reactive power flow from bus $i$ to $j$ \\
  $G_{ij}$ & equivalent conductance between bus $i$ and $j$\\
  $B_{ij}$ & equivalent susceptance between bus $i$ and $j$\\
  $\tau _{{p_i}}$ & active power time constant of the bus $i$ \\
  $\tau _{{q_i}}$ & reactive power time constant of the bus $i$ \\
\end{tabular}\\
\newline

This dynamic model and its similar types have been proposed in \cite{Nwankpa1992,Nguyen2015,Nguyen2016,Mohammed2000} to describe a variety of dynamic loads including {thermostatically controlled loads, induction motors, static loads, etc.} in ambient conditions, whereas the values of time constants depend on load types.  More importantly, the dynamic load model described above leads to a qualitatively correct load behavior over a wide range of voltage magnitudes \cite{DeMarco1990}. When load demand is taken as an independent input, the load states will respond to this input to maintain active and reactive power balance. According to (\ref{ddelta_dP})-(\ref{dV_dQ}), an increase in the load active power consumption (e.g., more negative $P_i^s$ in (\ref{ddelta_dP})) will lead to the bus voltage frequency drop; an increase in load reactive power consumption (e.g., more negative $Q_i^s$ in (\ref{dV_dQ})) will lead to the bus voltage frequency drop. 
Indeed, such behavior is qualitatively reasonable for many types of loads with different time constants. For example, the time constant for induction motor is in the range of few seconds, while the time constant of load tap changer is in the range of minutes. Even the static loads can be represented by applying the limit $\tau_{p_i}, \tau_{q_i} \to 0$. As a result, time constants can be selected to vary from cycles to minutes for different types of loads. In power systems, the load bus is the combination of many different types of loads, so the time constants of aggregated load in this paper lie in the range between 0.1s and 300s. 

Load powers are naturally experiencing fluctuations. In this paper, we make the common assumption, similar to %\color{red}cite also a paper not from our group  
\cite{Nwankpa1992,Mohammed2000,Pierrou2020},  % \sout{[10],[24]} \cite{Nwankpa1992,Mohammed2000}, 
that the load powers are perturbed by independent Gaussian variations from their static load power consumption:   
\
\small
\begin{eqnarray}
{\dot \delta _i} = \frac{1}{{{\tau _{{p_i}}}}}(P_i^s(1 + \sigma _i^p\xi _i^p) - {P_i})
\label{dynamic_load_angle_eqn}\\
{\dot V_i} = \frac{1}{{{\tau _{{q_i}}}}}(Q_i^s(1 + \sigma _i^q\xi _i^q) - {Q_i})
\label{dynamic_load_voltage_eqn}
\end{eqnarray}
\normalsize
where ${\xi _i^p}$ and ${\xi _i^q}$ are standard Gaussian random variables, ${\sigma _i^p}$ and ${\sigma _i^q}$ are the noise intensities of the load variations. %for static active and reactive power, respectively.

Assuming that the total number of buses inside the attacking region {in Fig. \ref{attacking_region} is $k$, by linearizing \eqref{dynamic_load_angle_eqn} and \eqref{dynamic_load_voltage_eqn} around the steady state, we have:}
\small
\begin{equation}
\label{swing-load-matrix}
\begin{array}{l}
\underbrace {\left[ \begin{array}{l}
{\bm{\dot \delta }}\\
{\bm{\dot V}}
\end{array} \right]}_{\bm{\dot x}} = \underbrace { - \left[ {\begin{array}{*{20}{c}}
{T_p^{ - 1}}&{}\\
{}&{T_q^{ - 1}}
\end{array}} \right]\underbrace {\left[ {\begin{array}{*{20}{c}}
{{J_{\bm{P\delta }}}}&{{J_{\bm{PV}}}}\\
{{J_{\bm{Q\delta }}}}&{{J_{\bm{QV}}}}
\end{array}} \right]}_J}_A\underbrace {\left[ \begin{array}{l}
\bm{\delta }\\
\bm{V}
\end{array} \right]}_{\bm{x}}\\
 + \underbrace {\left[ {\begin{array}{*{20}{c}}
{T_p^{ - 1}{P^s}{\Sigma ^p}}&0\\
0&{T_q^{ - 1}{Q^s}{\Sigma ^q}}
\end{array}} \right]}_B\underbrace {\left[ {\begin{array}{*{20}{c}}
{{\bm{\xi }^p}}\\
{{\bm{\xi }^q}}
\end{array}} \right]}_{\bm{\xi }}
\end{array}
\end{equation}
\normalsize
where $\bm{x}=[\bm{\delta}, \bm{V}]^T$;  $J_{\bm{P\delta}}=\frac{\partial\bm{P}}{\partial\bm{\delta}}$, $J_{\bm{PV}}=\frac{\partial\bm{P}}{\partial\bm{V}}$, $J_{\bm{Q\delta}}=\frac{\partial\bm{Q}}{\partial\bm{\delta}}$, $J_{\bm{QV}}=\frac{\partial\bm{Q}}{\partial\bm{V}}$ represent the Jacobian matrices.  

\begin{table}[ht]\normalsize
\begin{tabular}{ll}

$\bm{\delta } = {[{\delta _1},...,{\delta _t},...,{\delta _k}]^T}$, 
& ${T_p} = {\rm{diag}}[{\tau _{{p_1}}},...,{\tau _{{p_t}}},...,{\tau _{{p_k}}}]$,\\
$\bm{V} = {[{V_1},...,{V_t},...,{V_k}]^T}$,
& ${T_q} = {\rm{diag}}[{\tau _{{q_1}}},...,{\tau _{{q_t}}},...,{\tau _{{q_k}}}]$,\\
$\bm{P} = {[{P_1},...,{P_t},...,{P_k}]^T}$, 
& ${P^s} = {\rm{diag}}[P_1^s,...,P_t^s,...,P_k^s]$,\\
$\bm{Q} = {[{Q_1},...,{Q_t},...,{Q_k}]^T}$,
& ${Q^s} = {\rm{diag}}[Q_1^s,...,Q_t^s,...,Q_k^s]$,\\
${\bm{\xi }^p} = {[\xi _1^p,...,\xi _t^p,...,\xi _k^p]^T}$,
& ${\Sigma ^p} = {\rm{diag}}[\sigma _1^p,...,\sigma _t^p,...,\sigma _k^p]$, \\
${\bm{\xi }^q} = {[\xi _1^q,...,\xi _t^q,...,\xi _k^q]^T}$,
& ${\Sigma ^q} = {\rm{diag}}[\sigma _1^q,...,\sigma _t^q,...,\sigma _k^q]$, \\

\end{tabular}
% \caption{Notations for the power system dynamic model}
\label{notation}
\end{table}
Thus, the compact form of the dynamic load model can be represented %\textcolor{georgia}
{as a multivariate}
% by the stochastic differential equation, which is indeed 
Ornstein-Uhlenbeck process \cite{Gardiner2009}:
\small
\begin{equation}
\label{eq:OU}
\dot{\bm{x}}=A\bm{x}+B\bm{\xi}
\end{equation}
\normalsize
The system state matrix $A$, corresponding to a scaled Jacobian matrix, carries significant information about the time constants of the dynamic loads and the line parameters $G_{tj}$ and $B_{tj}$ inside the attacking region. In Section \ref{3d}, a purely measurement-based method is proposed to estimate the matrix $A$, {which allows the intruder to extract line parameters previously missing and to design the FDIA on AC state estimation based on (\ref{fake_P_ij})-(\ref{fake_I_i}). %\eqref{fake_P_tj}-\eqref{fake_Q_i} 

\subsection{Regression Theorem for the Ornstein-Uhlenbeck Process}
\label{3d}

The $\tau$-lag time correlation matrix $C(\tau)$ {of a stochastic process} in the stationary state is defined as:
\begin{equation}
C(\tau) = E\left( {\left( {\bm{x}(t + \tau ) - {\bm{\mu }_x}} \right){{\left( {\bm{x}(t) - {\bm{\mu }_x}} \right)}^T}} \right)
\label{covariance}
\end{equation}
where ${\bm{\mu}_x}$ denotes the mean of state variables ${\bm{x}}$. Note that $C(0)$ corresponds to the covariance matrix. When the power system operates in the normal steady state such that The system state matrix $A$ is stable, the regression theorem for the multivariate Ornstein-Uhlenbeck process states that the $\tau$-lag correlation matrix can be described by a differential equation \cite{Gardiner2009}:
\small
\begin{equation}
\frac{d}{d\tau}\left[C(\tau)\right]=AC(\tau)
\label{regressiontheorem}
\end{equation}
\normalsize

In other words, matrix $A$ can be obtained from the $\tau$-lag correlation matrix: 
\small
\begin{equation}
A =  \frac{1}{\tau }\ln [C(\tau )C{(0)^{ - 1}}]
\label{eq:estimation_of_matrix_A_tau}
\end{equation}
\normalsize
In practice, the intruder can only use  the sample mean ${\hat{\bm{\mu }}_x}$ and the sample correlation matrix $\hat{C}(\Delta t)$  to estimate ${\bm{\mu }_x}$ and $C(\tau)$ due to limited data:
\small
\begin{equation}
{\hat{\bm{\mu }}_x} = \frac{1}{N}\sum\limits_{i = 1}^N {{\bm{x}^{(i)}}}   \\
\label{mu_x}
\end{equation}

\begin{equation}
\hat{C}(0) = \frac{1}{{N - 1}}({F_{1:N}} - {\hat{\bm{\mu }}_x}{\bm{1}_{1:N}}){({F_{1:N}} - {\hat{\bm{\mu }}_x}{\bm{1}_{1:N}})^T} \\
\end{equation}
\label{C_zero}
\begin{equation}
\hat C(\Delta t) = \frac{1}{{N - 1}}({F_{1 + K:N}} - {\hat{\bm{\mu }}_x}{\bm{1}_{1:N - K}}){({F_{1:N - K}} - {\hat{\bm{\mu }}_x}{\bm{1}_{1:N - K}})^T}
\label{C_t}
\end{equation}
\normalsize
where $N$ is the sample size, $K$ is the number of samples that corresponds to the selected time lag $\Delta t$, ${\bm{x^{(i)}}} = {[x_1^{(i)},x_2^{(i)},...,x_M^{(i)}]^T}$ represents the ${i^{th}}$ measurement of $M$ state variables, i.e., the voltage angles and magnitudes, $F = [{x^{(1)}},{x^{(2)}},...,{x^{(N)}}]$ is a $M \times N$ matrix, $\bm{1}$ is a $1 \times N$ matrix of ones. Matrices ${F_{i:j}}$ and ${\bm{1}_{i:j}}$ denote the submatrices of $F$ and $\bm{1}$ from ${i^{th}}$ to ${j^{th}}$  column, respectively. 
{Applying (\ref{regressiontheorem}), matrix $A$ can be estimated from:}
\small
\begin{equation}
{\hat{A}} =  \frac{1}{\Delta t }\ln [\hat{C}(\Delta t )\hat{C}{(0)^{ - 1}}]
\label{eq:estimation_of_matrix_A_eqn}
\end{equation}
\normalsize

Equation (\ref{eq:estimation_of_matrix_A_eqn}) provides an intelligent way for the intruder to obtain the estimation of matrix $A$ purely from PMU measurements. The regression theorem has been utilized to estimate the dynamic system state matrix of power systems \cite{Sheng2020} and design wide-area damping control \cite{Guo2020}. Yet it is the first time that the theorem is applied for the study and design of FDIA against AC state estimation.  

In the next section, we will provide an estimation method to extract line parameters from the estimated matrix $\hat{A}$ for constructing the attack vector.

\subsection{Estimation for Time Constants and Line Parameters}
\label{3e}

To extract the Jacobian matrices and further line parameters from $\hat{A}$, the intruder needs to firstly estimate the time constants  $1/\tau_{p_i}, 1/\tau_{q_i}$ for all non boundary buses, i.e., $\forall i \in {\Omega _A}\setminus \Omega_B$.

To this end, %can apply the following 
the weighted least square estimation (WLS) optimization formulation \cite{Monticelli1999} can be applied.  

Generally, the WLS method can be used to estimate parameters $\bm{\beta}\in\mathbb{R}^{p}$ that satisfy $\bm{Y}=X\bm{\beta}$ by minimizing the residual:
\begin{equation}
\begin{aligned}
\arg \mathop {\min }\limits_{\bm{\beta }} \left\| {W(\bm{Y} - X\bm{\beta })} \right\|_2
\end{aligned}\label{eq:WLS}
\end{equation}
where $W\in\mathbb{R}^{n\times n}$ represents a weight matrix.  $\bm{Y}\in\mathbb{R}^n$ and $X\in\mathbb{R}^{n\times p}$ are obtained from measurement data of size $n$, $n>p$. If $X$ has full column rank, the solution to (\ref{eq:WLS}) is
\begin{equation}
\bm{\beta } = {({X^T}WX)^{ - 1}}{X^T}W\bm{Y}
\label{least_square_eqn_tau}
\end{equation}

For the estimation of $1/\tau_{p_i}, 1/\tau_{q_i}$,  $\bm{Y}$,  ${X}$ and $\bm{\beta}$ can be obtained from \eqref{dynamic_load_angle_eqn}} as follows:

\begin{equation}
\label{least_square_estimation_for_time_constant}
\underbrace {\left[ \begin{array}{l}
\frac{1}{{\Delta t}}(\delta _i^{(2)} - \delta _i^{(1)})\\
...\\
\frac{1}{{\Delta t}}(\delta _i^{(n)} - \delta _i^{(n - 1)})
\end{array} \right]}_{\bm{Y}} = \underbrace {\left[ \begin{array}{l}
{{\hat \mu }_{{P_i}}} - P_i^{(1)}\\
...\\
{{\hat \mu }_{{P_i}}} - P_i^{(n - 1)}
\end{array} \right]}_X\underbrace {\left[ {\frac{1}{{{\tau _{{p_i}}}}}} \right]}_{\bm{\beta}}
\end{equation}
where $P_i^{(n)}$  represents the  ${n^{th}}$  observation of active power of the  bus $i$ ; ${\hat{\mu} _{{P_i}}}$  denotes the sample mean of  $P_i$ , $\delta _i^{(n)}$  represents the   ${n^{th}}$  observation of voltage angle of  bus $i$ .

By substituting (\ref{least_square_estimation_for_time_constant}) into (\ref{least_square_eqn_tau}), the estimated  $1/{\hat{\tau} _{{p_i}}}$  can be obtained. Similarly, the intruder can calculate the estimated  $1/{\hat{\tau} _{{q_i}}}$  using \eqref{dynamic_load_voltage_eqn} through the observation of  $[Q_i^{(1)},Q_i^{(2)},...,Q_i^{(n)}], [V_i^{(1)},V_i^{(2)},...,V_i^{(n)}]$  with $n \ge 2$. Note that the power injections at bus $i$  can be calculated from the voltage and current phasors received from the PMU. 

Once the estimated time constants  ${\hat{\tau}_{{p_i}}},{\hat{\tau} _{{q_i}}}$  are obtained, the estimated derivatives $\hat{J}_{P_i\delta_j}$, ${{{\hat J}_{{P_i}{V_j}}}}$, $\hat{J}_{Q_i\delta_j}$, ${{{\hat J}_{{Q_i}{V_j}}}}$ , $\forall ( i,j)  \in {L_A}$ can be obtained from the scaled Jacobian matrices embedded in $\hat{A}$, which can be further utilized to estimate the line parameters  $G_{ij}$ and $B_{ij}$.

Taking derivatives for both sides of (\ref{bus active power injection}) and (\ref{bus reactive power injection}), one can obtain the following linear relationship between the Jacobian matrices, the measurements of voltage magnitudes and angles, and the line admittances: 
\small

\small
\begin{equation}
\label{least_square_estimation_for_G_B}
\underbrace {\left[ {\begin{array}{*{20}{c}}
{{{\hat J}_{{P_i}{\delta _j}}}}\\
{{{\hat J}_{{P_i}{V_j}}}}\\
{{{\hat J}_{{Q_i}{\delta _j}}}}\\
{{{\hat J}_{{Q_i}{V_j}}}}
\end{array}} \right]}_{\bm{Y}} = \underbrace {\left[ {\begin{array}{*{20}{c}}
{{V_i}{V_j}\sin {\delta _{ij}}}&{ - {V_i}{V_j}\cos {\delta _{ij}}}\\
{{V_i}\cos {\delta _{ij}}}&{{V_i}\sin {\delta _{ij}}}\\
{ - {V_i}{V_j}\cos {\delta _{ij}}}&{{V_i}{V_j}\sin {\delta _{ij}}}\\
{{V_i}\sin {\delta _{ij}}}&{ - {V_i}\cos {\delta _{ij}}}
\end{array}} \right]}_X\underbrace {\left[ {\begin{array}{*{20}{c}}
{{G_{ij}}}\\
{{B_{ij}}}
\end{array}} \right]}_{\bm{\beta}}
\end{equation}
\normalsize

By applying the WLS method  (\ref{least_square_eqn_tau}) again, the intruder can obtain the estimated line conductance  ${\hat{ G}_{ij}}$  and susceptance  ${\hat{B}_{ij}}$, $\forall (i,j) \in {L_A}$. Since $G_{ij}=G_{ji}$, an average of  ${\hat{ G}_{ij}}$ and  ${\hat{ G}_{ji}}$ can be used as an estimate of $G_{ij}$ and $G_{ji}$. 

Thus, a novel FDIA method can be developed, which requires no pre-knowledge of line parameters. The overall procedure for the {proposed FDIA method} is described in {Algorithm 1}.

\begin{algorithm}[ht]
\caption{FDIA without knowledge of line parameters.} 
\label{Flow chart of the FDIA} 
\begin{algorithmic}  %
\State
\textbf{Step 1}. Given one target bus $t$, define the attacking region  $S_A=(\Omega_A, L_A)$. Let $\Omega_B$ be the set of boundary buses.  \\ 
\textbf{Step 2}. $ \forall i \in {\Omega _A}$,  collect voltage  magnitude and angle measurements $V_i\angle \delta_i$ of size $N$ 
and current magnitude and angle measurements $I_i\angle \theta_i$  of size $n$ 
via intercepted PMUs. \\
\textbf{Step 3}. Calculate {the estimated matrix $\hat{A}$ through \eqref{mu_x}-\eqref{eq:estimation_of_matrix_A_eqn}}. \\
\textbf{Step 4}. Calculate the power injection  ${P_i},{Q_i}$ and estimate the time constants ${\hat{\tau}_{{p_i}}},{\hat{\tau}_{{q_i}}}$ $\forall i \in {\Omega _A}\setminus \Omega_B$,   %for the target bus $t$  
through \eqref{least_square_eqn_tau} and \eqref{least_square_estimation_for_time_constant}.\\ 
\textbf{Step 5}. Approximate the line admittance  {$\hat{G}_{ij},\hat{B}_{ij}$,  $\forall ( i,j)  \in {L_A}$, 
through \eqref{least_square_eqn_tau} and \eqref{least_square_estimation_for_G_B}. \\
\textbf{Step 6}. Given the target malicious phasor ${\tilde V_t}\angle {\tilde \delta _t}$, design the attack vector $\bm{a}$ through \eqref{fake_P_ij}-\eqref{fake_I_i}.\\ 
\textbf{Step 7}. Manipulate relevant RTU  and PMU measurements  according to $\bm{a}$ in \eqref{eq:attack vector}. 
}
\end{algorithmic}
\end{algorithm}

\textbf{Remarks:}\\
\noindent 1. %For the sample size $N$ and $n$ 
In \textbf{Step 2}, %in reality, 
$N$ needs to be sufficiently large to acquire accurate estimation for the stationary correlation matrix $\hat{C}(\Delta t )$ in \textbf{Step 3}, while $n$ can be small as only one parameter needs to be estimated in \eqref{least_square_estimation_for_time_constant} through WLS. 
In the simulations of this paper, $N=18000$ and $n=10$, i.e.,  window sizes of 300s and 0.167s with a frequency of 60Hz.

\noindent 2. Once the intruder collects the required voltage and angle measurements in \textbf{Step 2}, only a few milliseconds are needed to estimate the matrix and finally extract the line parameters and design the attack vector (\textbf{Step 3-Step 6}), indicating the high computational efficiency and the suitability of the proposed method for online implementation. The detailed computational time for the case studies is presented in Section \ref{sectionexampleI}. Besides, since the line parameters may vary, the online algorithm can estimate the line parameters within a 5 minute time window, ensuring the performance of the FDIA.

\noindent 3. The proposed FDIA against AC state estimation requires only limited PMU measurements inside the attacking region, while no other RTU  measurements (e.g., line flows) or line parameters are needed to construct the attack vector. 

\noindent 4. In \textbf{Step 6}, the selection of malicious target phasor ${\tilde V_t}\angle {\tilde \delta _t}$ is beyond the scope of the paper. 
It can be selected in light of specific purposes of the intruder (e.g., affecting the real-time electricity price\cite{Jia2014},  resulting in power outages\cite{Liang2017}) while ensuring that the fundamental limits known to the control room are not violated \cite{Liu2017}. 
{One advantage of the proposed FDIA model is that targeted large deviation attacks can be launched as will be shown in Section \ref{casestudy}}. 

\noindent 5. The data manipulation in \textbf{Step 7} may happen on the local RTU/PMU equipment side, during the data transmission in wide area network (WAN), or on the remote control center side.

\subsection{Sufficient Conditions for the Proposed FDIA Model}

Sufficient conditions for the proposed FDIA model against AC state estimation are provided in this section. Although the conditions may be strong in practice, they provide important insights to the performance of the proposed method. In addition, it will be shown in Section \ref{casestudy} that the proposed FDIA model can bypass the BDD with a very high probability even under large targeted deviation attacks in practical applications. 

As discussed in  \cite{Rahman2013}, \cite{Liu2017}, 
it is challenging to analytically investigate the change of residual due to the complexity of the nonlinear model.
%as discussed in \cite{Rahman2013}\cite{Liu2017}. 
Recall the standard steps for solving the AC state estimation problem:
\begin{enumerate}
    \item[] 1: Set $\Delta  \bm{x}=(H^{T}H)^{-1}H^T(\bm{z}-\bm{h(x)})$
    \item[] 2: $\bm{x}_{k+1}=\bm{x}_k+\Delta \bm{x}$
    \item[] 3: update $H$
    \item[] 4: update $\bm{h(x)}$. 
\end{enumerate}
The above steps are repeated until the state vector converges to a fixed point. %
Besides, if Dishonest Gauss Newton method is applied, the Jacobian matrix $H$ will remain fixed through iterations such that Step 3 can be omitted. 
{Note that the convergence of Gauss/Dishonest Gauss Newton algorithms is not analytically guaranteed and the exact fixed point cannot be obtained analytically. }

Assuming that without the attack, the solution of the applied numerical algorithm converges to the true state values. Thus, we have that if $\Delta \bm{x} \rightarrow 0$, then $T(\bm{z}-\bm{h(\hat{x})})=0$, where $T=(H^{T}H)^{-1}H^T$ is the pseudo inverse of $H$, then we have the following theorem, providing sufficient conditions under which the proposed FDIA model is perfect.

\noindent \textit{Theorem 1}. If AC state estimation is exploited using numerical methods that do not update the Jacobian matrix $H$ in each iteration (e.g., Dishonest Gauss Newton method), and the following conditions are satisfied: {(i)} the attack vector is sufficiently small and thus can be represented linearly; 
{(ii)} $H$ is a square matrix and $rank(H)$ is equal to the number of states $\bm{x}$ to be estimated, then the FDIA designed in {Algorithm 1} is perfect, i.e., the residual after the attack remains the same as that before the attack.

\noindent \textit{Proof:} Let $\hat{\bm{h}}$ denote the nonlinear functions obtained at the intruder side using the estimated line admittances $\hat{G}_{tj}$ and $\hat{B}_{tj}$ in \textbf{Step 5}, then with condition (i), we have $\bm{a}=\hat H\bm{c}$, where $\hat H$ is the estimated Jacobian matrix from $\hat{\bm{h}}$, and thus $\bm{z}_{bad}=\bm{z}+\bm{a}=\bm{z}+\hat H\bm{c}$. Assuming that the numerical algorithm converges to a fixed point close to the true state values $\hat{\bm{x}}_{bad}=\hat{\bm{x}}+\hat{\bm{c}}$, where $\hat{\bm{c}}$ is small. Then if $\Delta \bm{x} \rightarrow 0$ %$\Delta \bm{x} \rightarrow 0$ 
in the numerical algorithm, we have:

\begin{eqnarray}
&&T({\bm{z}_{bad}} - \bm{h}({{\bm{\hat x}}_{bad}})) = 0\nonumber\\
&\Rightarrow &T(\bm{z} + \hat H\bm{c} - \bm{h}({{\bm{\hat x}}_{bad}})) = 0\nonumber\\
&\Rightarrow &T(\bm{z} + \hat H\bm{c} - (\bm{h}(\bm{\hat x}) + H\bm{\hat c}) = 0\nonumber\\
&\Rightarrow &T(\bm{z} - \bm{h}(\bm{\hat x})) + T(\hat H\bm{c} - H\bm{\hat c}) = 0\nonumber\\
&\Rightarrow &T(\hat H\bm{c} - H\bm{\hat c}) = 0\nonumber\\
&\Rightarrow &{\bm{\hat c}} = T\hat{H}\bm{c}\label{eq:ACmodel_c_hat}
\end{eqnarray}
note that $\hat{\bm{c}}$ is typically different from $\bm{c}$ intended, because of the estimation error in $\hat{\bm{h}}$. 

Therefore, the residual after the designed attack in Algorithm 1 can be calculated:
\begin{eqnarray}
{\bm{r}_{bad}}& =& \|{\bm{z}_{bad} - {\bm{h}}(\bm{\hat x}}_{bad})\|_\infty\nonumber\\
& = &\|\bm{z} + \hat H\bm{c}\nonumber- ({\bm{h}}(\hat{\bm{x}})+{H}\hat{\bm{c}})\|_\infty\nonumber\\
& = &\|\bm{z}- {\bm{h}}(\hat{\bm{x}})+\hat{H}\bm{c}-{H}\hat{\bm{c}}\|_\infty\nonumber\\
%& = &\|\bm{z}-{\bm{h}}(\hat{\bm{x}}) + (I - HT)\hat{H}\bm{c}\|_\infty\nonumber\\
&=&\|\bm{z}-{\bm{h}}(\hat{\bm{x}})  + (I - H{({H^T}H)^{ - 1}}{H^T})\hat{H}\bm{c}\|_\infty\label{eq:ACmodel}
\end{eqnarray}
with condition (ii), $H$ is invertible, (\ref{eq:ACmodel}) becomes $r_{bad}=\|\bm{z}-{\bm{h}}(\hat{\bm{x}})\|=r$. This completes the proof.

Note that it is assumed that the fixed point that the numerical algorithm converges to satisfies  $\hat{\bm{x}}_{bad}=\bm{\hat{x}}+\bm{\hat{c}}$, where $\hat{\bm{x}}$ denotes the estimated states before the attack. Although it cannot be theoretically guaranteed, it has been shown in \cite{Rahman2013} that it is typically true in practical state estimation scenarios.

In addition, it is worth mentioning that although the conditions may be too strict for practical applications, by applying \textit{Theorem 1} we can achieve by far the best analytical results. In case the conditions are relaxed, e.g., if the attack vector is large such that condition (i) is not satisfied,  or the Jacobian matrix is not a constant but is updated at each iteration,  then the attack cannot be guaranteed to be perfect. I.e., the residual may change and the attack is not 100\% successful. Similar difficulty exists in previous works \cite{Hug2012,Rahman2013,Liu2017}. Since there is no analytical method to determine the residual of the AC state estimation, the residual should rather be calculated by numerical simulation. However, compared to previous works that rely on the linear model $\bm{a}=H\bm{c}$ \cite{Yuan2011a,Rahman2012,Deng2019}, the proposed attack vector is designed using the nonlinear model $\bm{a}=\bm{h}(\bm{\hat{x}}+\bm{c})-\bm{h}(\bm{\hat{x}})$. As will be shown in Section IV, the proposed method can still be used to launch large deviation attacks with a high success rate (but not 100\%).  Besides, as the number of measurements is typically greater than that of states, %
%In practice, 
condition (ii) may also be violated in practice so that the residual after the attack is not guaranteed to remain unchanged. Nevertheless, thanks to the accurate estimation result for  the line parameters and thus $\bm{h}$ %$H$ 
by the proposed algorithm, it will be shown in Section \ref{casestudy} that the launching of large deviation attacks may not result in big residuals, indicating a high probability of the proposed FDIA model to bypass the BDD in practical applications.

\normalsize

\section{Case Studies}\label{casestudy}

In this section, the proposed FDIA model against AC state estimation is implemented on a modified IEEE 39-bus 10-generator system. 

We consider the worst situation from the perspective of the intruder, i.e, the power system is fully measured by RTUs for state estimation at the control room, while the intruder can only collect limited PMU data within the attacking region.  Therefore, each bus has two RTU measurements for active and reactive power, respectively; each line has four RTU measurements for active and reactive powers of two directions, resulting in a total of 262 RTU measurements. All the 262 RTU measurements (78 for bus power injections and 184 for line flows) are contaminated by independent Gaussian noises with standard deviation 0.05, i.e., $\forall e_i \in \bm{e}$ , ${e_i}\sim {\rm{ }}\mathcal{N}(0,{0.05^2})$ in (\ref{relationship between z and x}). It should be noted that all RTU measurements are used by the control room for state estimation and are not exploited by the intruder for designing attack vectors.  In addition, PMUs may also be utilized for state estimation at the control room. The results of the proposed FDIA model in the case where PMU data is used for the state estimation will be discussed in Section \ref{subsection:PMUstateestimation}.  

Besides, all the 10 generators are modeled by the second-order dynamic model, i.e., the swing equations \cite{Kundur1993}, while the remaining 29 buses are modeled as the dynamic loads described by (\ref{ddelta_dP})-(\ref{bus reactive power injection}). For the zero-injection buses, the time constants are selected to be 0.1. For the load buses, the time constants are drawn from a uniform distribution from $[0.5, 300]$, as the time constants of aggregated loads range from seconds to minutes.

All simulations have been carried out in Matlab using %the FDIA process is coded in Matlab and performed on 
a computer with Intel(R) Core(TM) i7-9750H CPU @ 2.60 GHz processor and 16.0 GB RAM.

\subsection{Constructing the proposed FDIA without pre-knowledge of line parameters}\label{sectionexampleI}

\begin{figure}[htbp]
\centering
\includegraphics[width=0.35\textwidth,keepaspectratio=true,angle=0]{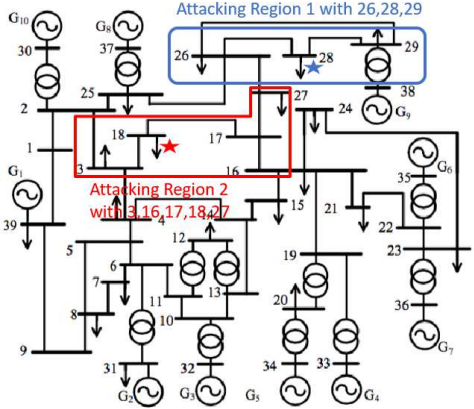}
\caption{The attacking region in IEEE 39-bus system.}
\label{False data attack in IEEE39 system}
\end{figure}

{Assuming the intruder intends to manipulate the states of bus 28, then the attacking region (Attacking Region 1 in Fig. \ref{False data attack in IEEE39 system}) includes bus 28 as well as buses 26 and 29 that are directly connected to bus 28, as per the definition in Section  \ref{section_attackingregion}.} 
Next, we will show that the intruder can stealthily bypass BDD to manipulate $V_{28}\angle {\delta_{28}}$  through the proposed {Algorithm 1}. As mentioned before, all the non-generator buses are modeled as dynamic load buses.  The process noise intensities $\sigma_i^p$, $\sigma_i^q$ of the 29 dynamic buses in \eqref{dynamic_load_angle_eqn} and \eqref{dynamic_load_voltage_eqn} describing power variations are set to be 1.  Particularly, the time constants for the loads inside the attacking region are $[{\tau _{{p_{26}}}},{\tau _{{p_{28}}}},{\tau _{{p_{29}}}}] =[45.83,221.32,204.16]$ s, $[{\tau _{{q_{26}}}},{\tau _{{q_{28}}}},{\tau _{{q_{29}}}}] = [{\rm{60}}{\rm{.62,12}}{\rm{.81,92}}{\rm{.38}}]$ s. PMU  measurement  noise  is  also  considered  to  test  the  performance  of  the measurement-based algorithm in a realistic setup. Independent Gaussian noise with zero mean and $10\%$ standard deviation of the largest state changes have been added to intruder's PMU measurements in \textbf{Step 2} similar to \cite{Zhou2013}.

According to {Algorithm 1}, the intruder needs to intercept and collect only voltage phasor measurements for all the buses inside the attacking region (i.e., bus 28, bus 26, and bus 29) as well as the current phasor of the target bus 28  from PMUs installed at bus 28, 26 and 29.  Particularly, 300s PMU measurements with a sampling rate of 60Hz, i.e., $N=18000$ samples, are used in \textbf{Step 2} to get the estimated matrix $\hat{A}$ in \textbf{Step 3}, after which the time constants and the line admittances can be estimated in \textbf{Step 4} and \textbf{Step 5}, respectively. As shown in Table \ref{Regression Theorem for OU process}, both the time constants in seconds (s) and the line admittances in per unit (pu) can be accurately estimated by the proposed algorithm. 

\begin{figure} [htbp]
\centering
\subfigure[Voltage of bus 26]
{ 
\label{bar1a} 
\begin{minipage}[t]{0.45\linewidth}
\centering
\includegraphics[width=1.5in]{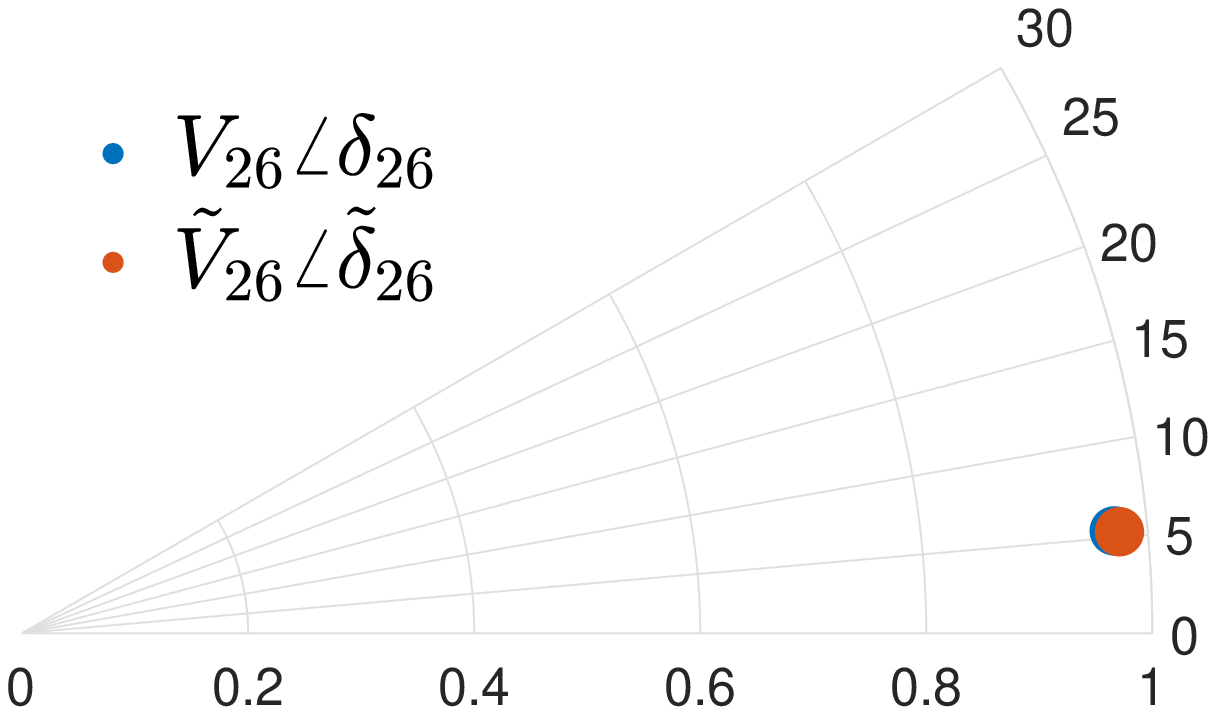}
\end{minipage}%
}
\subfigure[Voltage of bus 28]
{ 
\label{bar1b} 
\begin{minipage}[t]{0.45\linewidth}
\centering
\includegraphics[width=1.5in]{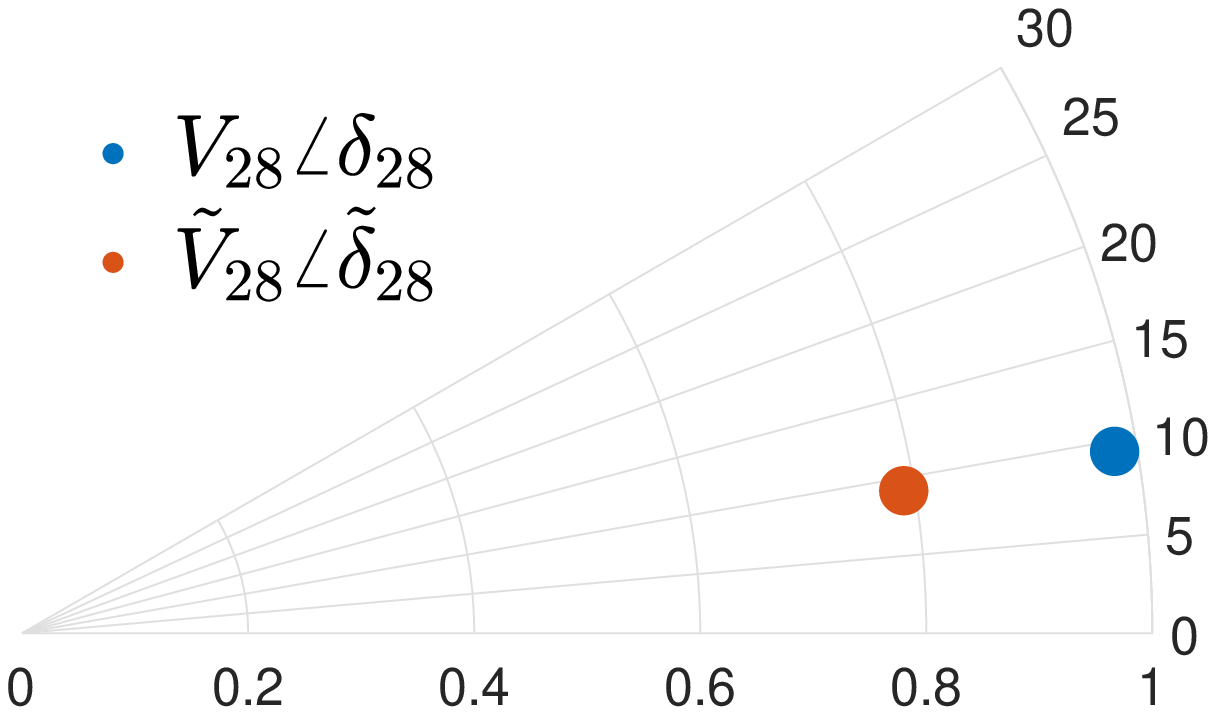}
\end{minipage}%
}%
\subfigure[Voltage of bus 29]
{ 
\label{bar1c} 
\begin{minipage}[t]{0.45\linewidth}
\centering
\includegraphics[width=1.5in]{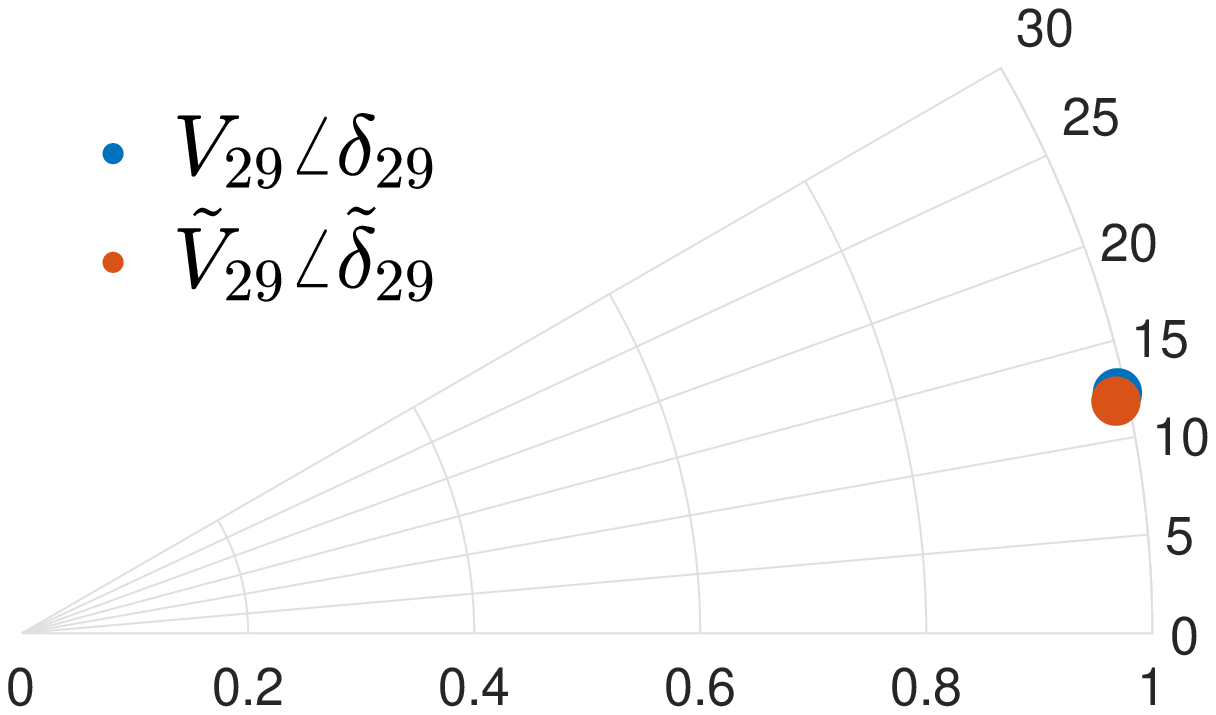}
\end{minipage}%
}%
\centering
\caption{False data attack with ${\tilde V_{28}}=0.8V_{28}$, ${\tilde \delta_{28}}= \delta_{28}$  against the WLS state estimator. }
\label{voltage magnitude and angle before and after attack_region1} 
\end{figure}

Once the line admittances are estimated, the intruder can design the attack vector $\bm{a}$ in \textbf{Step 6} according to (\ref{eq:attack vector}). Assuming that the attacker intends to launch an attack on the voltage magnitude $\tilde V_{28}=0.8{V_{28}}$, Fig. \ref{voltage magnitude and angle before and after attack_region1} presents a comparison between the estimated state values through the AC  WLS state estimation \cite{Monticelli1999} at the control room before and after the attack. The radial coordinate represents the voltage magnitude (in pu) and the angular coordinate represents the voltage angles (in degrees). {It can be observed that the attack is launched successfully as designed. Meanwhile, the residual after the attack $r_{bad}=0.5188$ has a negligible difference compared to that before the attack $r=0.5117$, indicating that the attack can successfully bypass BDD.}

\normalsize
\begin{table}[!th]
\centering
\captionsetup{justification=centering}
  \caption{\textsc{True and estimated values of line parameters and time constants }}\label{Regression Theorem for OU process}
\begin{tabular}{|c|c|c|}
\hline
\textbf{Time constant}& \textbf{True value (s)}& \textbf{Estimated value (s)}\\
 \hline
$\tau_{p_{28}}$&221.32&220.12
\\
\hline
$\tau_{q_{28}}$&12.81&12.86
\\
\hline
\textbf{Line Parameter}& \textbf{True value (pu)}& \textbf{Estimated value (pu)}\\
\hline
\begin{tabular}{c} ${G_{26 - 28}} + j{B_{26 - 28}}$ \end{tabular}&1.898-20.928i &1.960-20.899i
\\
\hline
\begin{tabular}{c} ${G_{29 - 28}} + j{B_{29 - 28}}$ \end{tabular}&	6.087-65.661i &	6.053-61.467i
\\
\hline
  \end{tabular}
  \
\end{table}

\normalsize

\begin{figure} [htbp]
\centering
\subfigure[]
{ 
\begin{minipage}[t]{0.5\linewidth}
\centering
\includegraphics[width=1.8in]{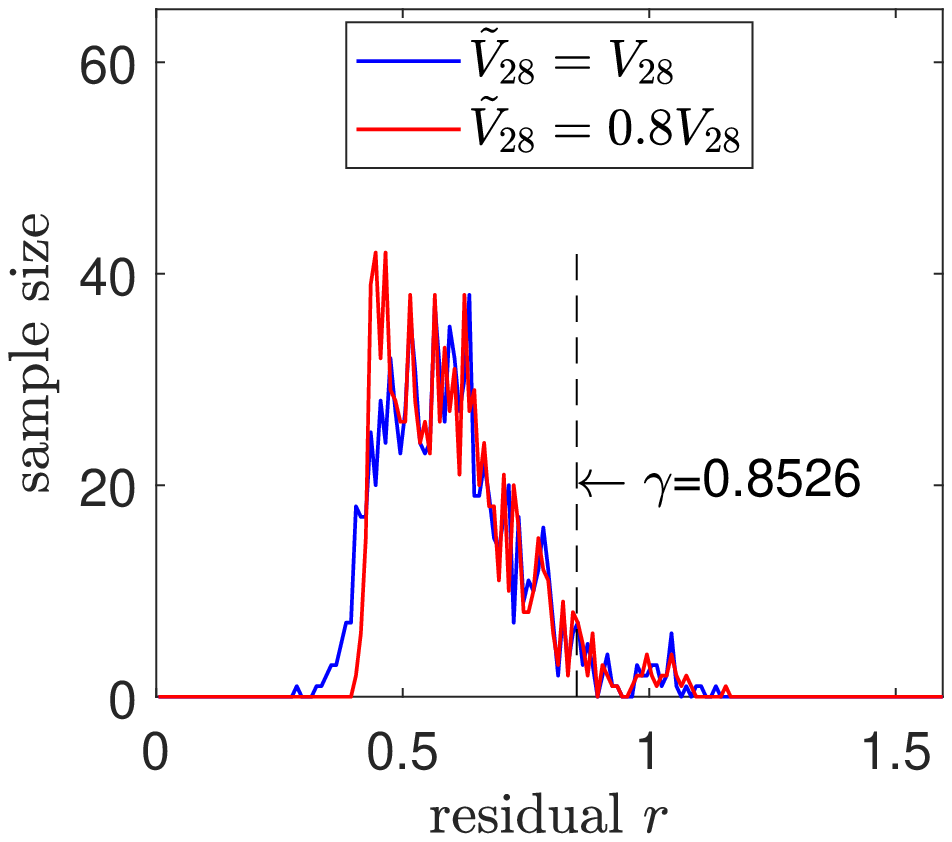}
\end{minipage}%
}%
\subfigure[]
{ 
\begin{minipage}[t]{0.5\linewidth}
\centering
\includegraphics[width=1.8in]{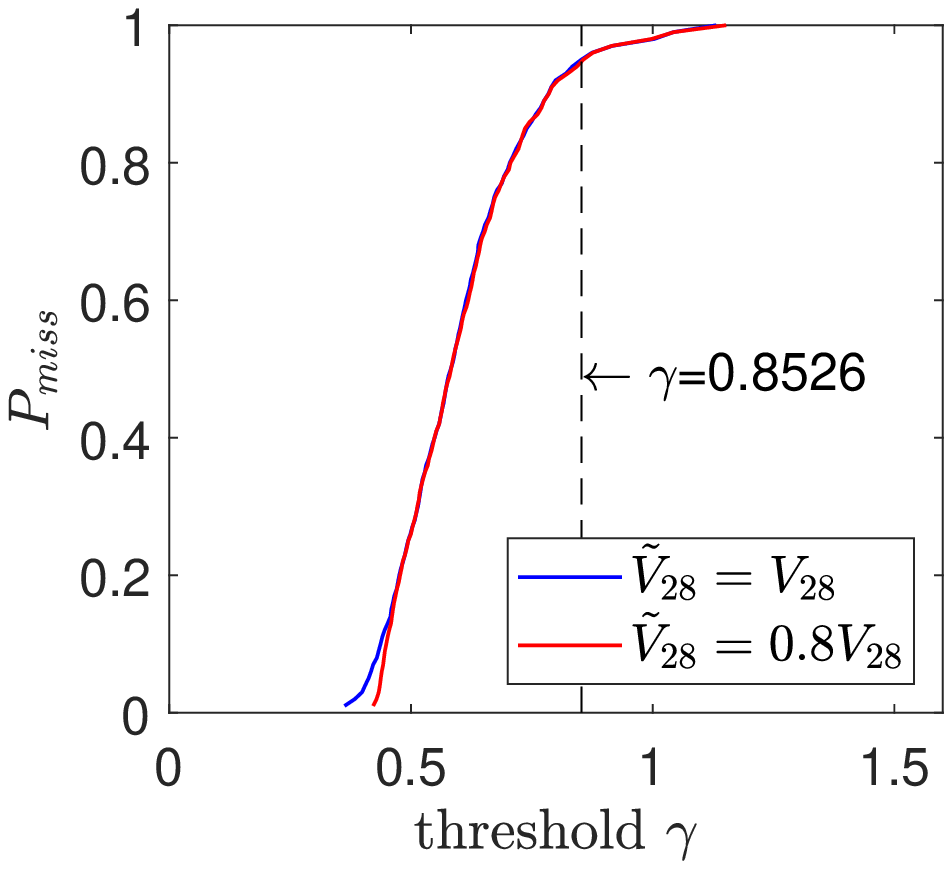}
\end{minipage}%
}
\centering
\caption{(a). The residual distribution before and after the attack $\tilde V_{28}=0.8{V_{28}}$ under the WLS state estimator; (b). The probability to bypass the WLS estimator. } 
\label{residual_distribution_WLS} 
\end{figure}

To further demonstrate the feasibility of the proposed FDIA model when the WLS state estimation is implemented at the control room, 1000 Monte Carlo simulations are run, in which the same attack $\tilde V_{28}=0.8{V_{28}}$ is implemented.  As shown in Fig. \ref{residual_distribution_WLS}, the designed attack does not result in significant changes in the distribution of the residual of the WLS state estimator.

In addition, if we choose the $95\%$-quantile $\gamma=0.8526$ to be the bad data detection threshold, the attacks can successfully bypass the  WLS state estimator and launch the designed attack with a probability of 94.8\%.  Note that the $95\%$-quantile $\gamma=0.8526$ threshold is selected from residual distribution of the base case (i.e., without FDIA). The average time needed for designing one FDIA vector is $0.0043$ s (\textbf{Step 3-Step 6} in Algorithm 1).  

\subsection{Impacts of Different Attack Vectors}\label{section_diffattacks} % and Sample 
In this section, different targeted attacks will be investigated to demonstrate the effectiveness of the proposed FDIA model against AC state estimation, despite the estimation error of line admittances $G_{tj}$ and $B_{tj}$.  
With the same parameters in Section \ref{sectionexampleI}, various Monte Carlo simulations using 1000 samples have been carried out for different targeted attacks.  The same threshold as that in Section \ref{sectionexampleI} for the BDD, i.e., $95\%$-quantile $\gamma=0.8526$ from the base case, is applied. \color{black}

\begin{figure} [htbp]
\centering
\subfigure[]
{ 
\begin{minipage}[t]{0.5\linewidth}
\centering
\includegraphics[width=1.8in]{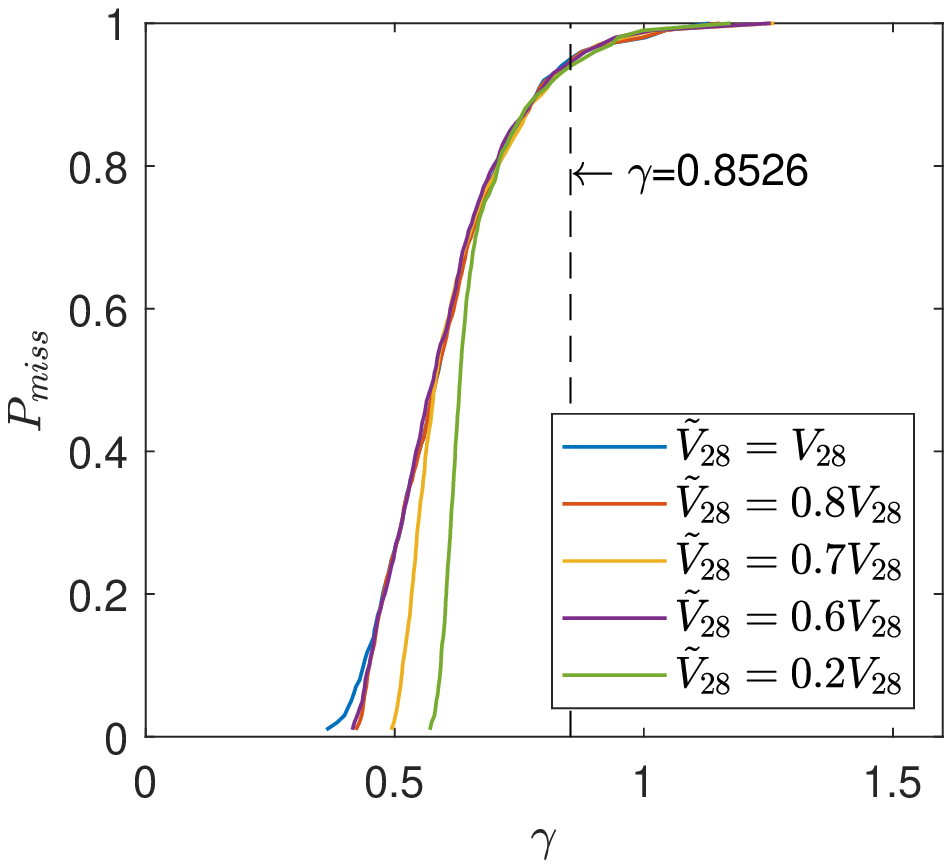}
\end{minipage}%
}%
\subfigure[]
{ 
\begin{minipage}[t]{0.5\linewidth}
\centering
\includegraphics[width=1.8in]{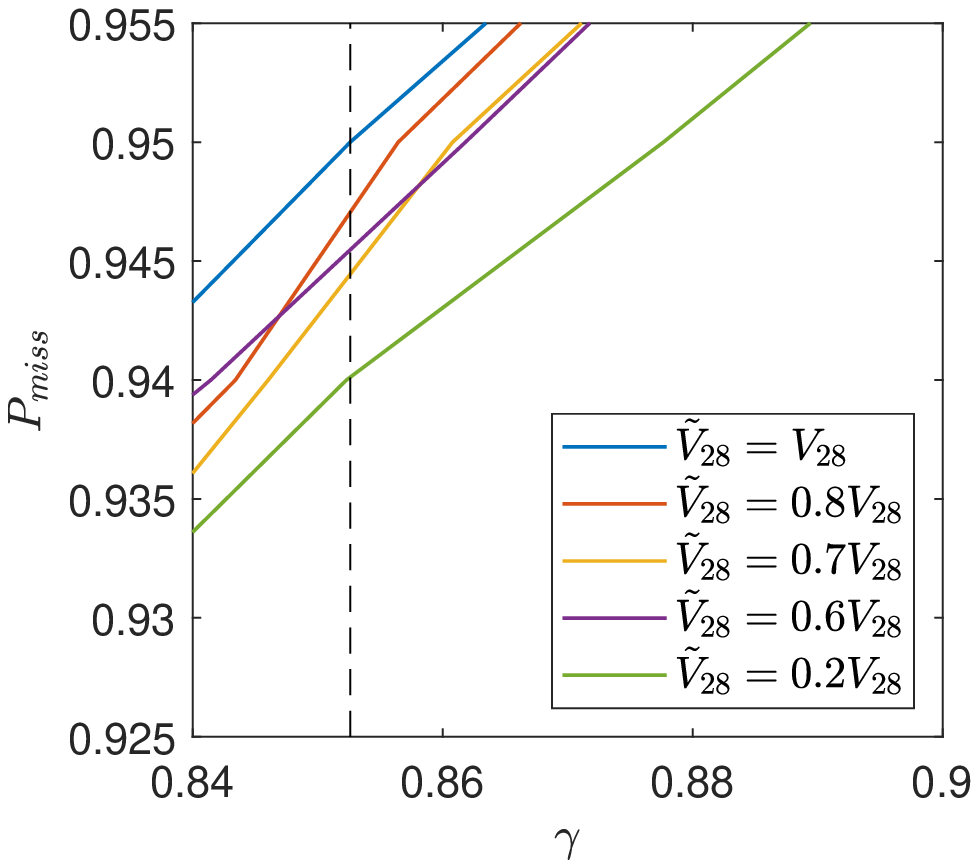}
\end{minipage}%
}%
\centering
\caption{(a). Probability to pass the WLS state estimator under different voltage magnitude attacks; (b). Zoomed-in picture of (a). }%Regional enlarged of (a). } 
\label{attack for V} 
\end{figure}

Fig. \ref{attack for V} presents the probabilities of the FDIAs to bypass the  WLS state estimator  with different voltage magnitude attacks, while keeping the voltage angle unchanged. The probabilities to bypass the BDD of ${\tilde V_{28}} = 0.8{V_{28}}$, ${\tilde V_{28}} = 0.7{V_{28}}$, ${\tilde V_{28}} = 0.6{V_{28}}$,  ${\tilde V_{28}} = 0.2{V_{28}}$ are 94.8\%, 94.4\%, 94.5\% and 94.1\%, respectively, showing that the intruder may successfully bypass the BDD when launching an up to {80\%} deviation attack on voltage magnitude.

\begin{figure} [htbp]
\centering
\subfigure[]
{ 
\begin{minipage}[t]{0.5\linewidth}
\centering
\includegraphics[width=1.8in]{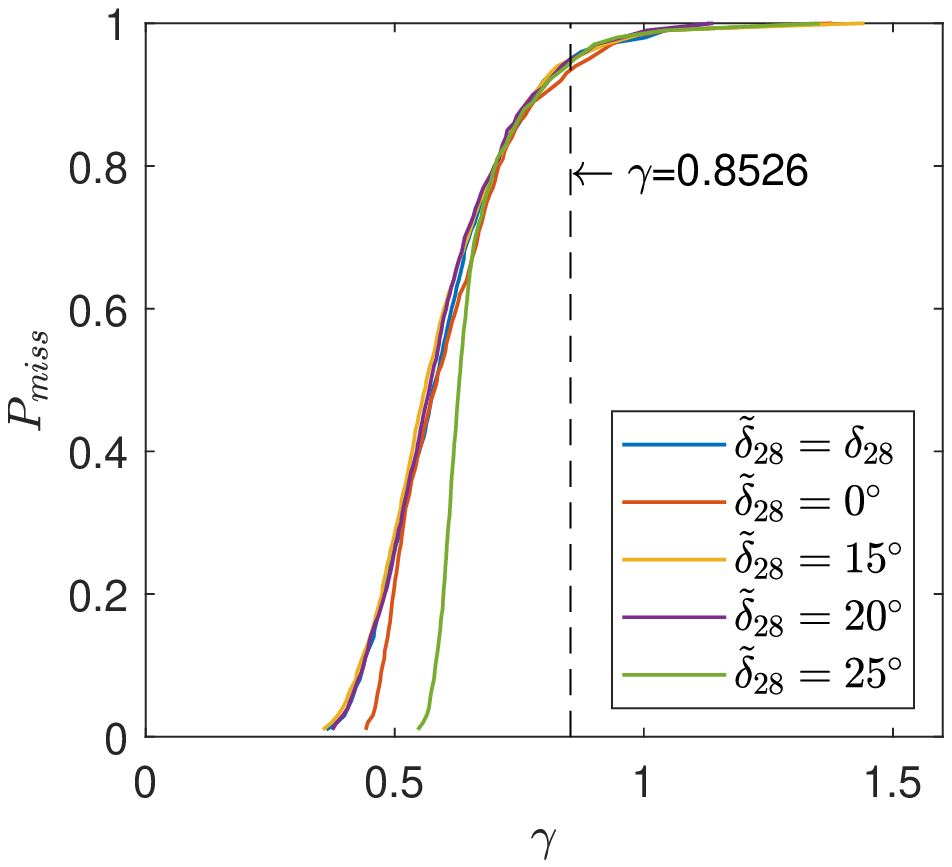}
\end{minipage}%
}%
\subfigure[]
{ 
\begin{minipage}[t]{0.5\linewidth}
\centering
\includegraphics[width=1.8in]{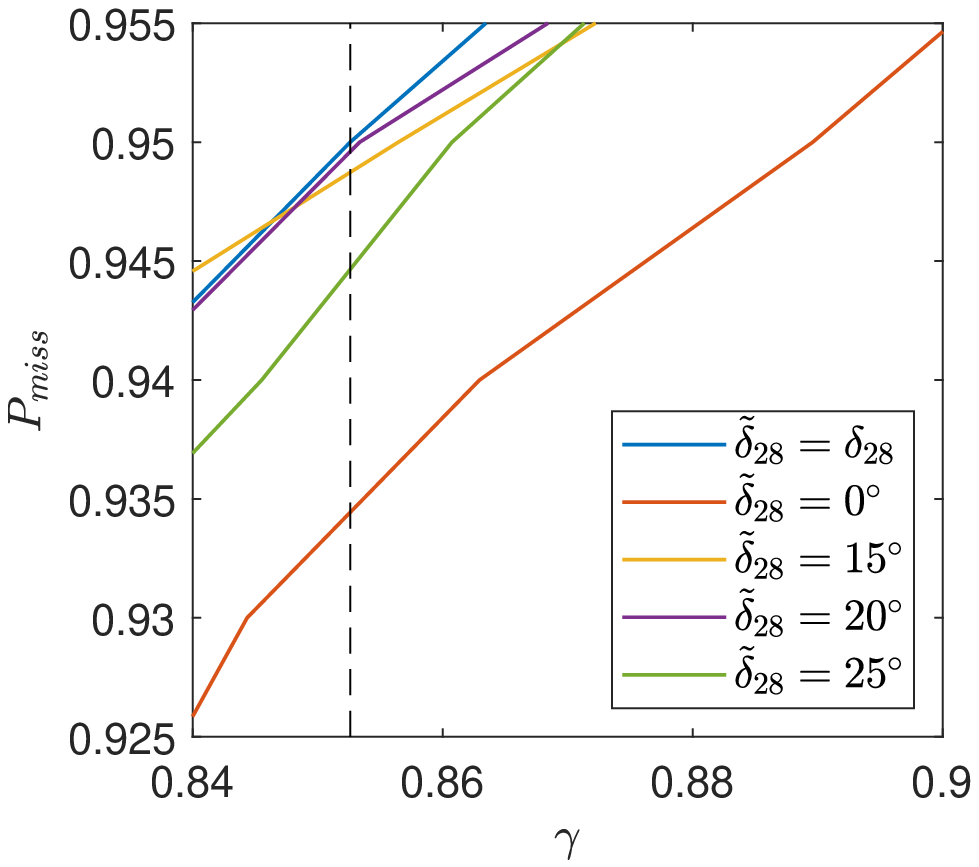}
\end{minipage}%
}%
\centering
\caption{(a). Probability to pass the  WLS state estimator  under different voltage angle attacks; (b). Zoomed-in picture of (a). }%Regional enlarged of (a). } 
\label{attack for delta} 
\end{figure}

Next, attacks on the voltage angle are considered. Normally,  ${V_{28}}\angle {\delta _{28}} = 0.98\angle {9.33^ \circ }$. It is assumed that the intruder intends to launch attacks such that ${\tilde \delta _{28}} = {0^ \circ }$, ${\tilde \delta _{28}} = {15^ \circ }$, ${\tilde \delta _{28}} = {20^ \circ }$, ${\tilde \delta _{28}} = {25^ \circ }$ respectively, while keeping the voltage magnitudes the same, i.e. ${\tilde V _{28}} = {V _{28}}$. 
As shown in Fig. \ref{attack for delta}, the probabilities to bypass the  WLS state estimator  for the aforementioned attacks are 93.5\%, 94.9\%, 95.0\%, and 94.5\%, respectively.

\begin{figure} [htbp]
\centering
\subfigure[]
{ 
\begin{minipage}[t]{0.5\linewidth}
\centering
\includegraphics[width=1.8in]{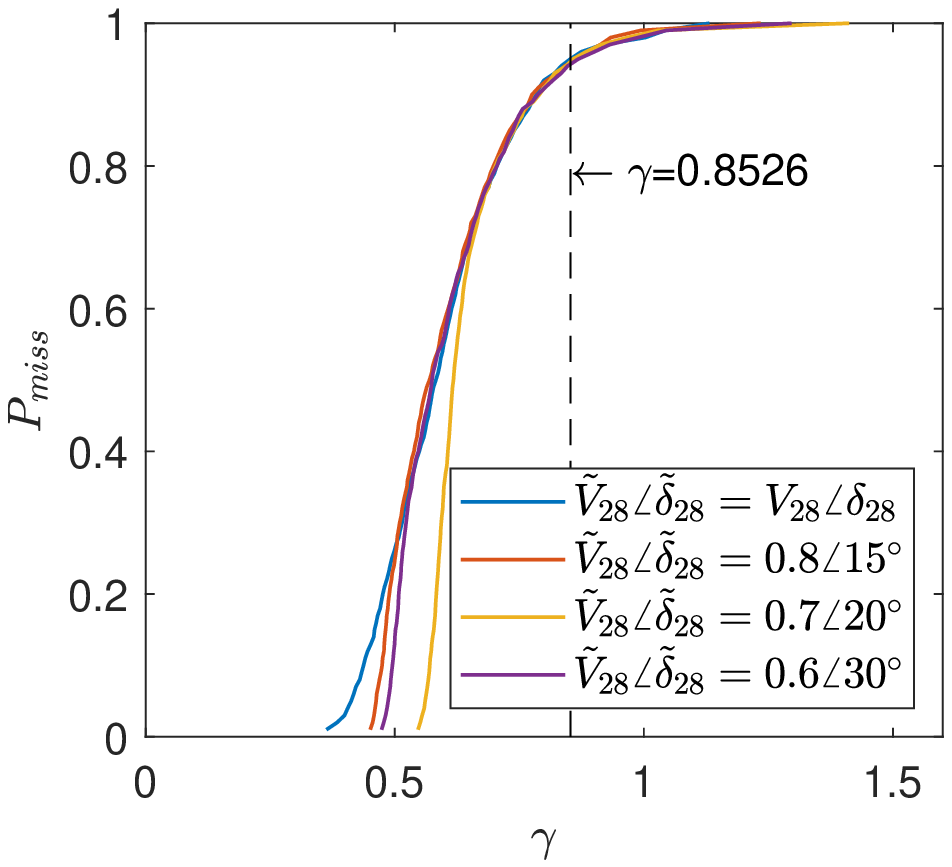}
\end{minipage}%
}%
\subfigure[]
{ 
\begin{minipage}[t]{0.5\linewidth}
\centering
\includegraphics[width=1.8in]{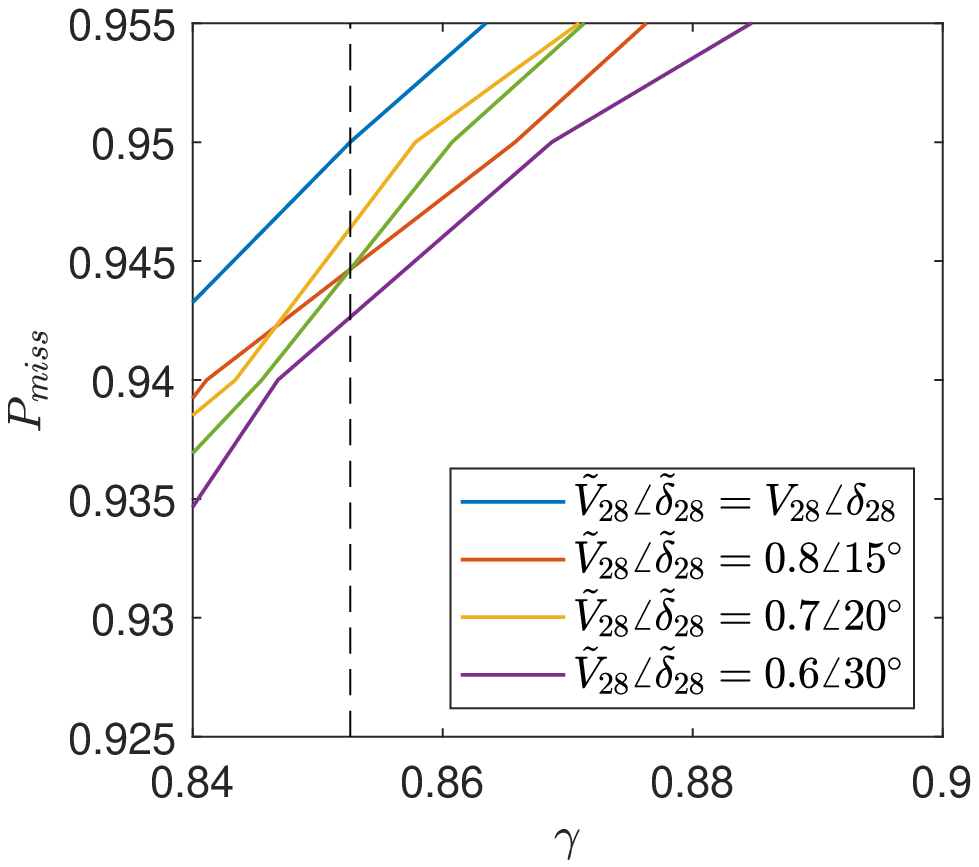}
\end{minipage}%
}%
\centering
\caption{(a). Probability to pass the  WLS state estimator  for both voltage magnitude and angle attacks; (b). Zoomed-in picture of (a). } %Regional enlarged of (a). } 
\label{attack for V and delta} 
\end{figure}

Lastly, attacks on both voltage magnitude and angle are studied. 
Fig. \ref{attack for V and delta} shows that the probabilities to bypass the  WLS state estimator  with $\tilde V_{28} \angle \tilde \delta_{28} = 0.8\angle {15^ \circ }$, $0.7\angle {20^ \circ }$, $0.6\angle {30^ \circ }$ are 94.3\%, 94.6\%, and 94.2\%, respectively, demonstrating that the proposed FDIA model can launch targeted large deviation attacks with high probabilities.

\subsection{Impacts of Different Attacking Regions}\label{subsectionexampleII} 
In this section, a different and larger attacking region is considered.  Particularly, a zero-injection bus is encountered when defining the attacking region.  Assuming the intruder intends to attack the states of bus 18,  then bus 3 and bus 17 will be included in the attacking region as per the definition. Yet bus 17 is a zero-injection bus, so bus 16 and bus 27 have to be included into the attacking region.  Eventually, the attacking region (Attacking Region 2 in Fig.  \ref{False data attack in IEEE39 system}) includes buses  3, 16, 17, 18 and 27. All buses inside the attacking region are considered to be dynamic load buses with time constants $[{\tau _{{p_3}}},{\tau _{{p_{16}}}},{\tau _{{p_{17}}}},{\tau _{{p_{18}}}},{\tau _{{p_{27}}}}] = {\rm{[181,20}}{\rm{.6,0.1,16}}{\rm{.1,28]}}$ s, $[{\tau _{{q_3}}},{\tau _{{q_{16}}}},{\tau _{{q_{17}}}},{\tau _{{q_{18}}}},{\tau _{{q_{27}}}}] = {\rm{[41,28}}{\rm{.5,0.1,23}}{\rm{.7,17}}{\rm{.5]}}$ s. The process noise intensities $\sigma_i^p$, $\sigma_i^q$, and PMU measurement noises are the same as those in Section \ref{sectionexampleI}.

Before the attacks, ${V_{18}}\angle {\delta _{18}} = 0.958\angle {\rm{2}}{\rm{.2}}{{\rm{1}}^ \circ },{V_{17}}\angle {\delta _{17}} = {\rm{0}}{\rm{.959}}\angle {\rm{3}}{\rm{.1}}{{\rm{9}}^ \circ }$.  Due to the existence of the zero injection bus 17, the intruder needs to alter both the states of bus 18 (the target bus) and that of bus 17 to ensure that the total power injection of bus 17 is zero, while the states of the other buses (i.e., boundary buses) inside the attacking region are not affected. Utilizing the constraint that the power injection of bus 17 is zero, it can be calculated that ${{\tilde V}_{17}}\angle {{\tilde \delta }_{17}} = {\rm{0}}{\rm{.893}}\angle {\rm{2}}{\rm{.4}}{{\rm{8}}^ \circ }$. 

\begin{figure} [htbp]
\centering
\subfigure[Voltage of bus 3]
{ 

\begin{minipage}[t]{0.45\linewidth}
\centering
\includegraphics[width=1.5in]{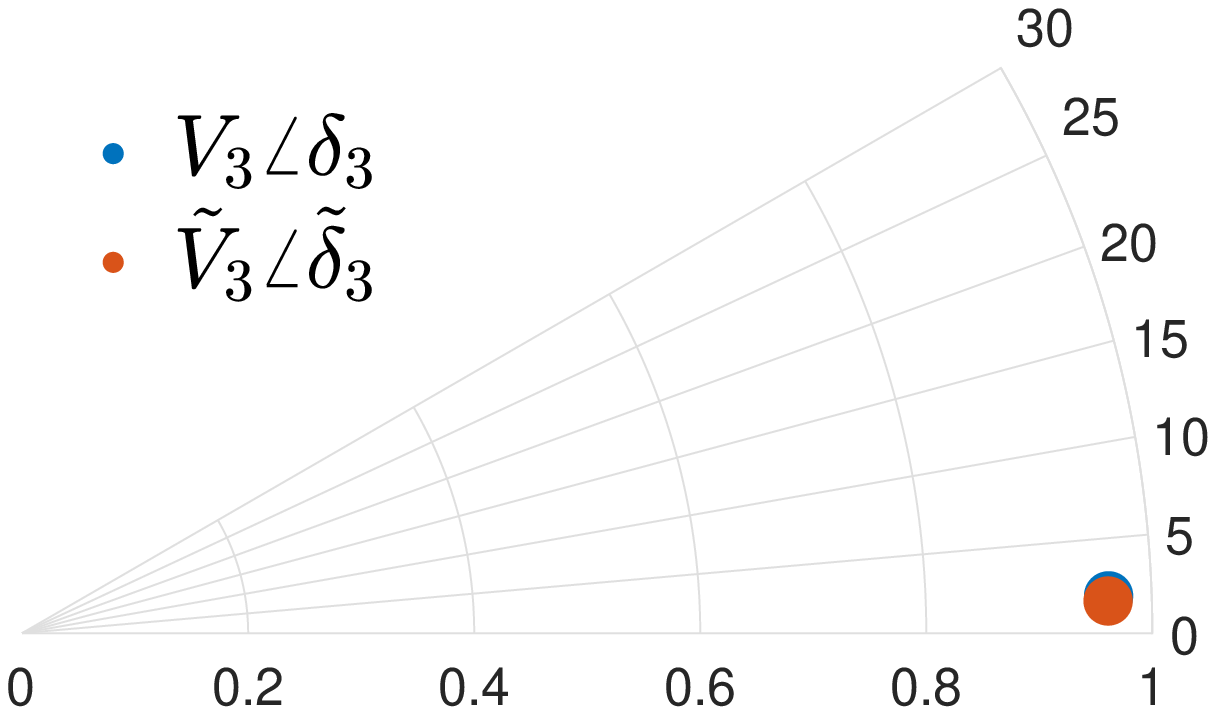}
\end{minipage}%
}
\subfigure[Voltage of bus 16]
{ 

\begin{minipage}[t]{0.45\linewidth}
\centering
\includegraphics[width=1.5in]{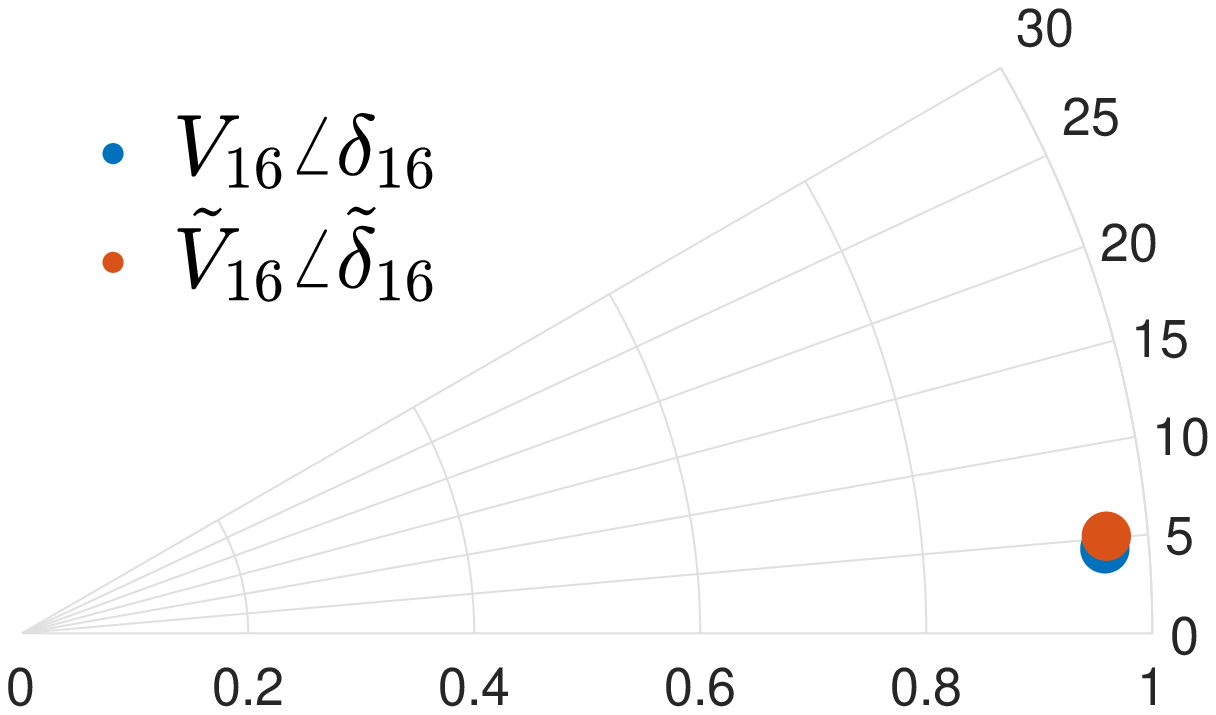}
\end{minipage}%
}%
\subfigure[Voltage of bus 17]
{ 
 
\begin{minipage}[t]{0.45\linewidth}
\centering
\includegraphics[width=1.5in]{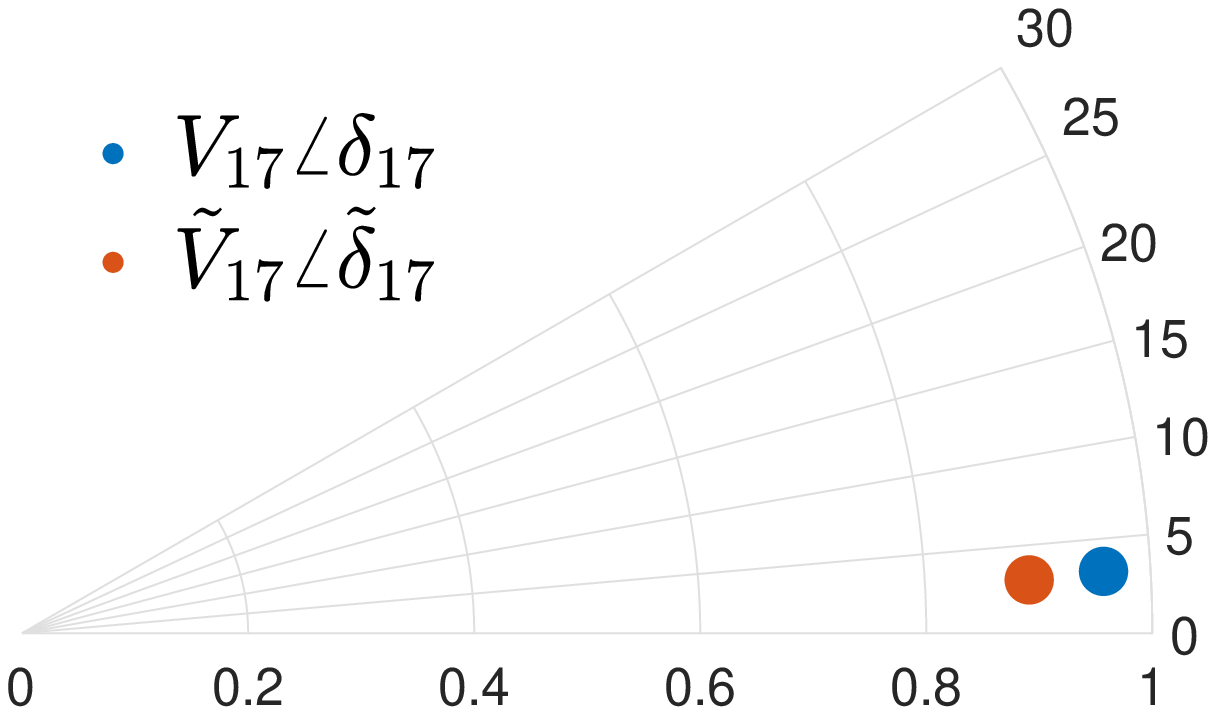}
\end{minipage}%
}%
\\
\subfigure[Voltage of bus 18]
{ 

\begin{minipage}[t]{0.45\linewidth}
\centering
\includegraphics[width=1.5in]{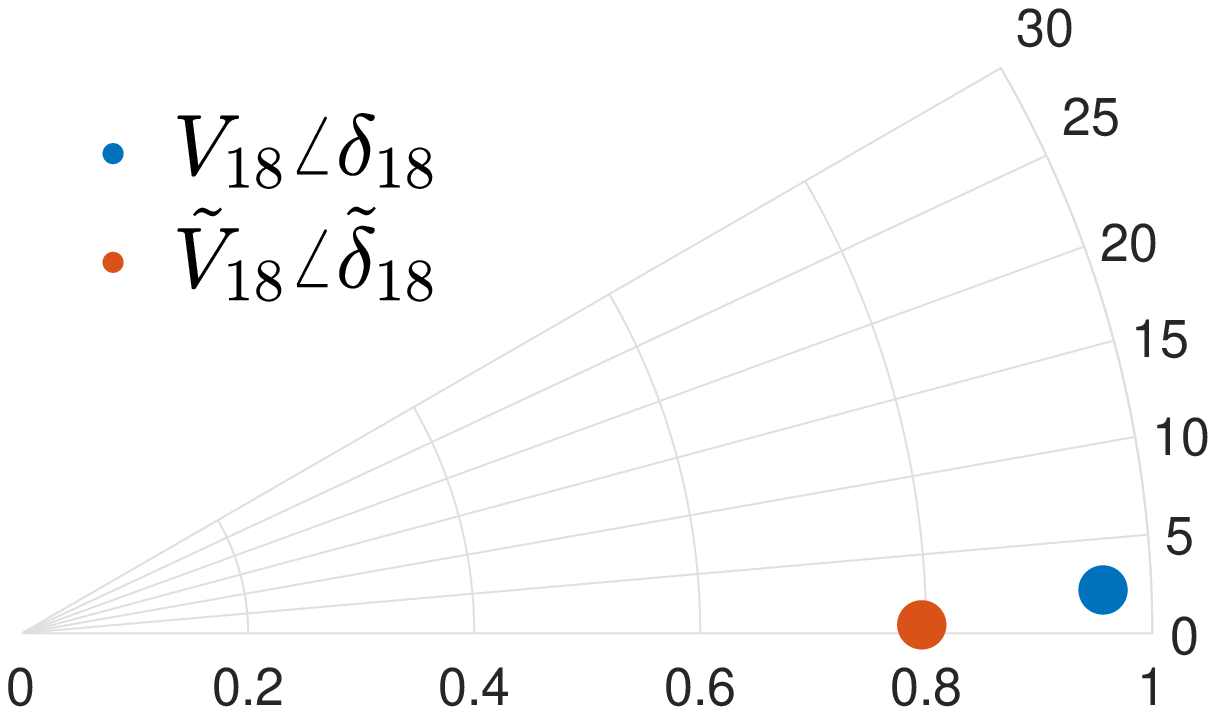}
\end{minipage}%
}%
\subfigure[Voltage of bus 27]
{ 

\begin{minipage}[t]{0.45\linewidth}
\centering
\includegraphics[width=1.5in]{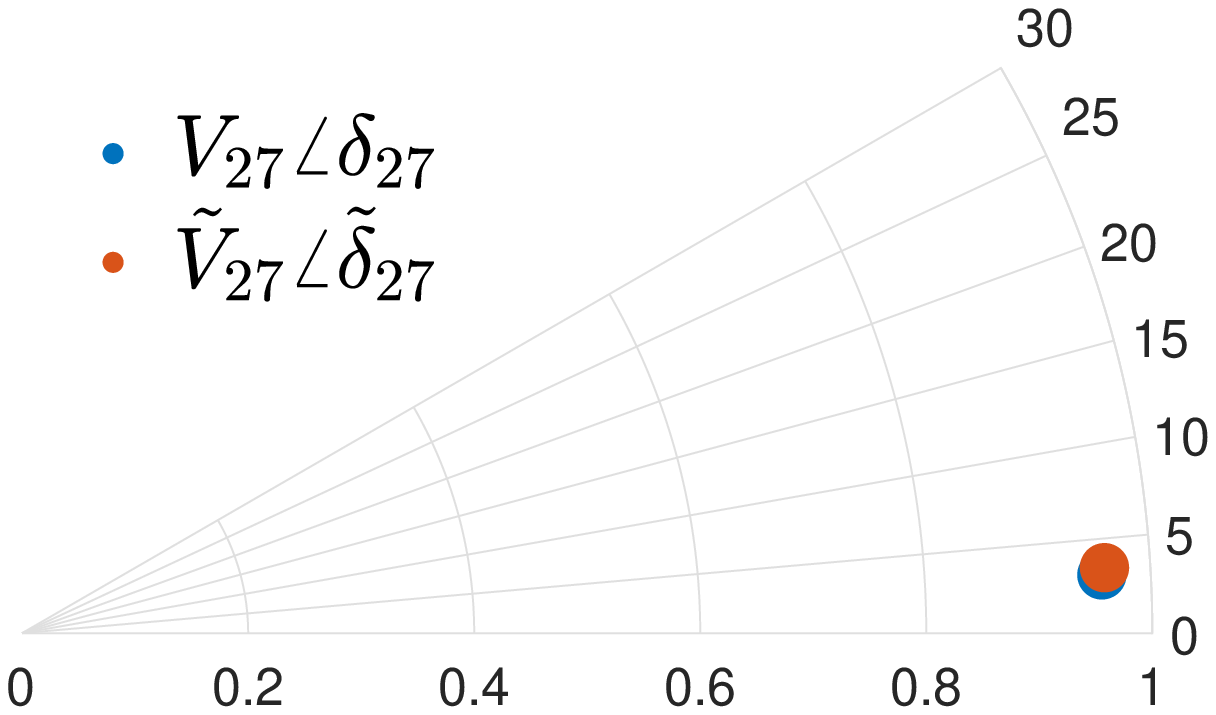}
\end{minipage}%
}%
\centering
\caption{False data attack with ${\tilde V_{18}}=0.8$, ${\tilde \delta_{18}}= 0$ against the WLS state estimator.}
\label{attack_region_2} 
\end{figure}

As observed from Fig. \ref{attack_region_2}, the attack can be launched successfully as designed. Although the voltage phasor of the zero-injection bus 17 has to be changed after the FDIA to ensure a zero power injection, the residual after the attack $r_{bad}=0.4468$ has a negligible difference compared to that before the attack $r=0.4403$, indicating that the attack can successfully bypass the BDD.

\begin{figure} [htbp]
\centering
\subfigure[]
{ 
\begin{minipage}[t]{0.5\linewidth}
\centering
\includegraphics[width=1.8in]{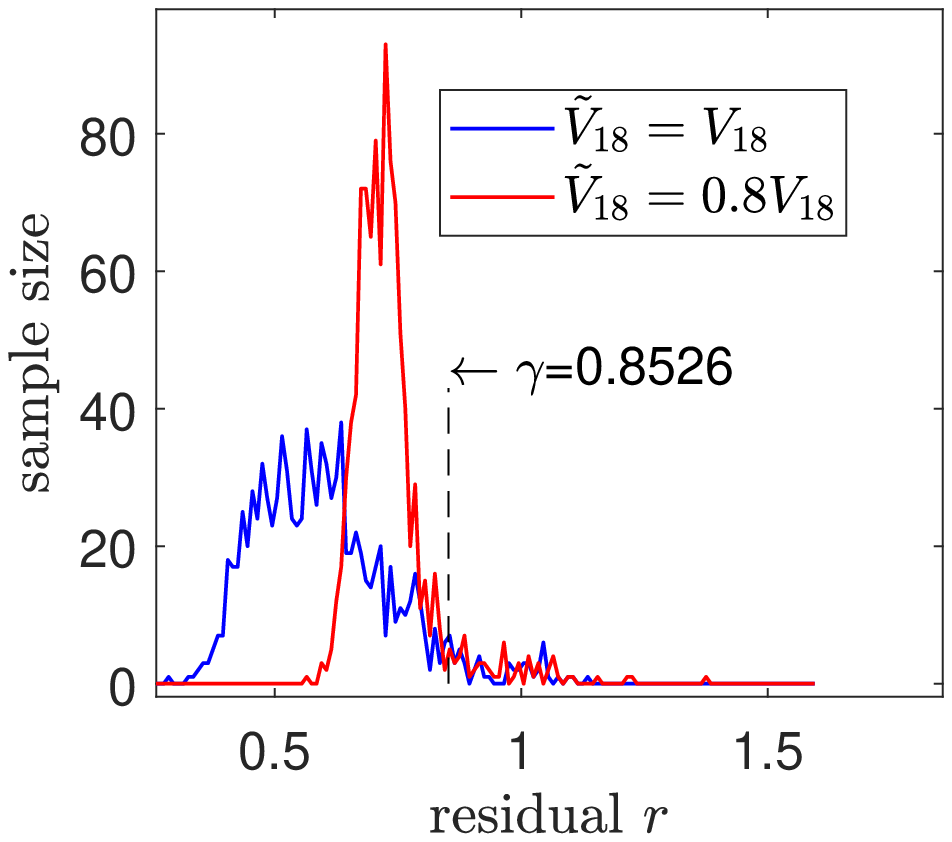}
\end{minipage}%
}%
\subfigure[]
{ 
\begin{minipage}[t]{0.5\linewidth}
\centering
\includegraphics[width=1.8in]{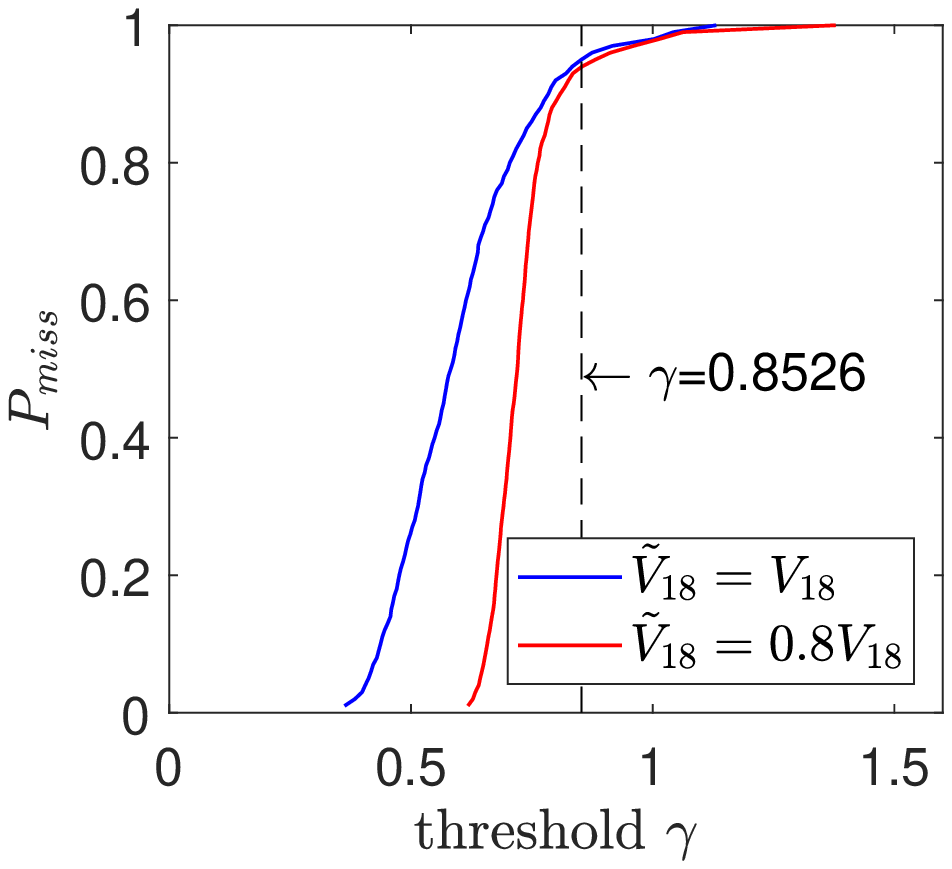}
\end{minipage}%
}
\centering
\caption{(a). The residual distribution before and after the attack $\tilde V_{18}=0.8{V_{18}}$ under WLS estimator in Attacking Region 2; (b). The probability to bypass the WLS estimator in Attacking Region 2. } 
\label{different_attacking_region}
\end{figure}

Likewise, 1000 Monte Carlo simulations are run, in which the attack $\tilde V_{18}=0.8{V_{18}}$ is implemented. As shown in Fig. \ref{different_attacking_region}, although the designed attack slightly changes the distribution of the residual compared to Fig. \ref{residual_distribution_WLS}, the probabilities to bypass the BDD of %${\tilde V_{28}} = 0.8{V_{28}}$  
${\tilde V_{18}} = 0.8{V_{18}}$ is still relatively high, i.e., 93.7\%.  Note that the same threshold for the BDD as that in Section \ref{sectionexampleI} is applied, i.e., $95\%$-quantile $\gamma=0.8526$ from the residual distribution of the base case. \color{black}

Due to the existence of the zero-injection bus, the intruder needs to expand the attacking region, which means more measurements need to be tampered with. Meanwhile, a new voltage equilibrium should be redesigned not only for the target bus but also for the zero-injection bus. These two reasons may result in a slight decrease in the success rate of the proposed FDIA in Attacking Region 2.

\subsection{Impacts of Different Load Power Perturbation Levels}
In practice, the stochastic load perturbation levels may vary in a wide range from time to time. In this section, the performance of the proposed FDIA method is tested under different stochastic load perturbation levels. 

We take Attacking Region 1 as an example when the target bus is bus 28.  
%Still focusing on Attacking Region 1 when the target bus is bus 28, 
For comparison, we launch the same FDIA, i.e., $\tilde V_{28}=0.8{V_{28}},\tilde \delta_{28}=\delta_{28}$, as in Section \ref{section_attackingregion}, 
%I.e, $\tilde V_{28}=0.8{V_{28}},\tilde \delta_{28}=\delta_{28}$ 
while considering different load power perturbation intensities, i.e., $\sigma _i^p = \sigma _i^q = 10,\sigma _i^p = \sigma _i^q = 1,\sigma _i^p = \sigma _i^q = 0.1,\sigma _i^p = \sigma _i^q = 0.01$ for all the 29 dynamic loads.

\begin{figure} [htbp]
\centering
\subfigure[]
{ 
\begin{minipage}[t]{0.5\linewidth}
\centering
\includegraphics[width=1.8in]{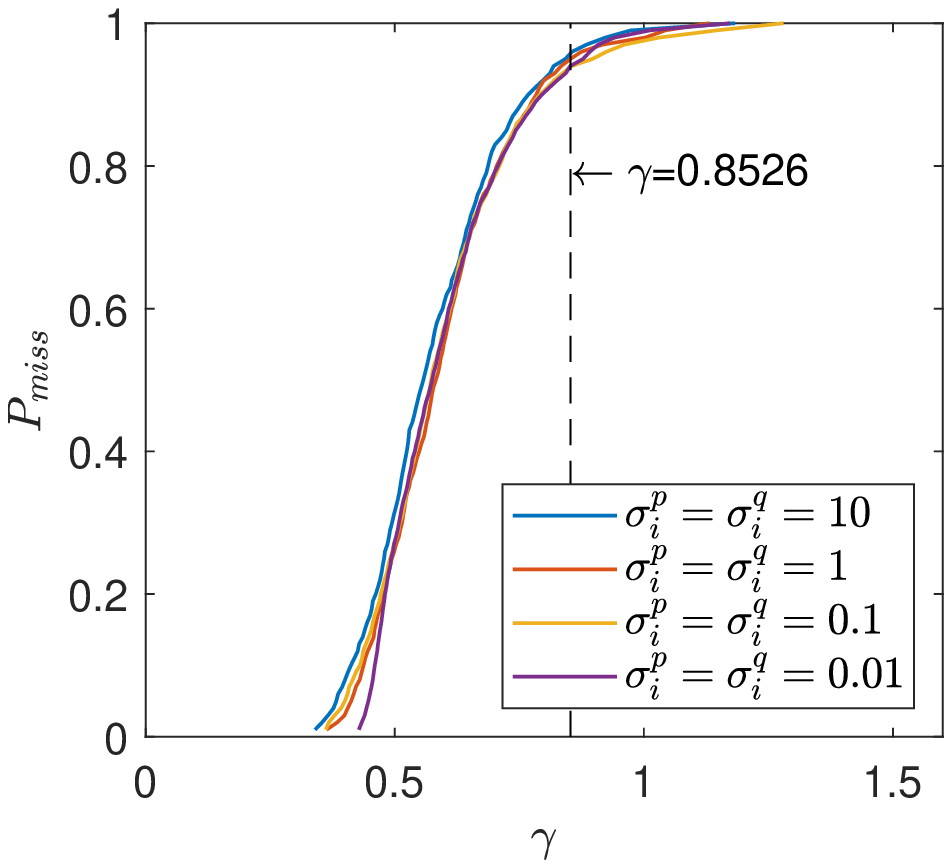}
\end{minipage}%
}%
\subfigure[]
{ 
\begin{minipage}[t]{0.5\linewidth}
\centering
\includegraphics[width=1.8in]{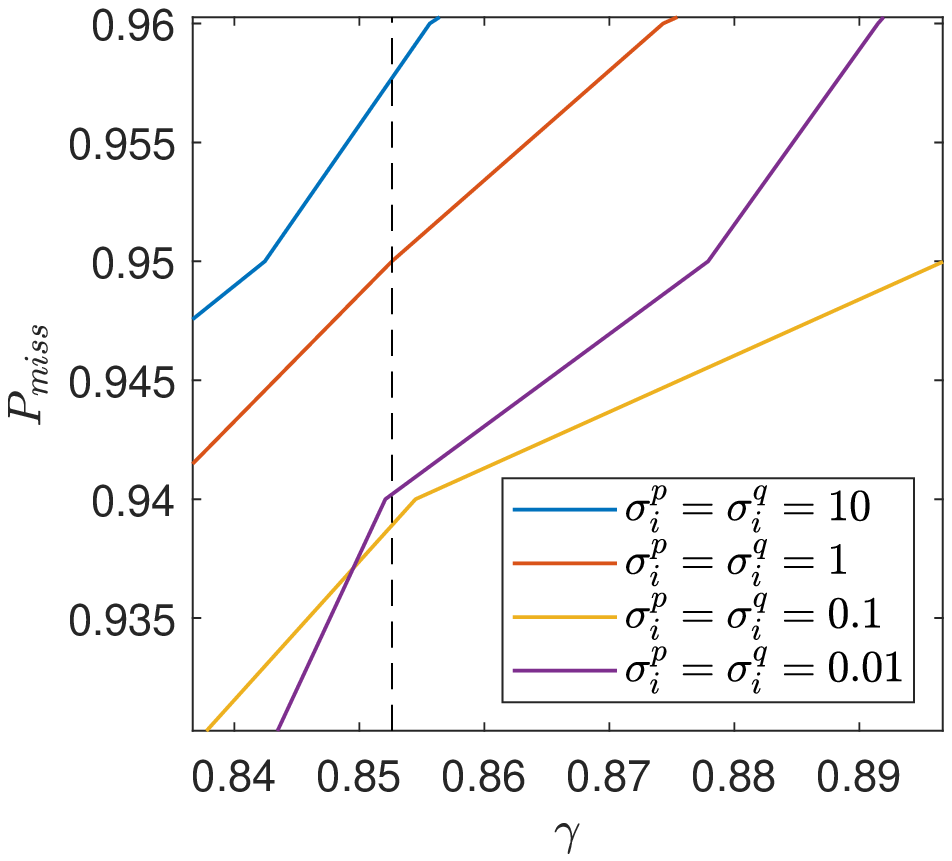}
\end{minipage}%
}
\centering
\caption{(a). Probability to pass the WLS state estimator using different load variations; (b). Zoomed-in picture of (a). }
\label{different_load_pertubation}
\end{figure}
As shown in Fig. \ref{different_load_pertubation}, the probabilities of the proposed FDIA to bypass the  WLS state estimator for the aforementioned attack under different load power perturbation levels are 95.7\%, 95.0\%, 94.0\%, and 93.8\%, respectively, showing that different load perturbation levels will not significantly deteriorate the performance of the proposed FDIA model.  The same threshold for the BDD as that in Section \ref{sectionexampleI} is applied, i.e., $95\%$-quantile $\gamma=0.8526$ from the residual distribution of the base case. \color{black}

\subsection{Impacts of Different PMU Sample Sizes}\label{section:sample size}
\begin{figure} [htbp]
\centering
\subfigure[]
{ 
\begin{minipage}[t]{0.5\linewidth}
\centering
\includegraphics[width=1.8in]{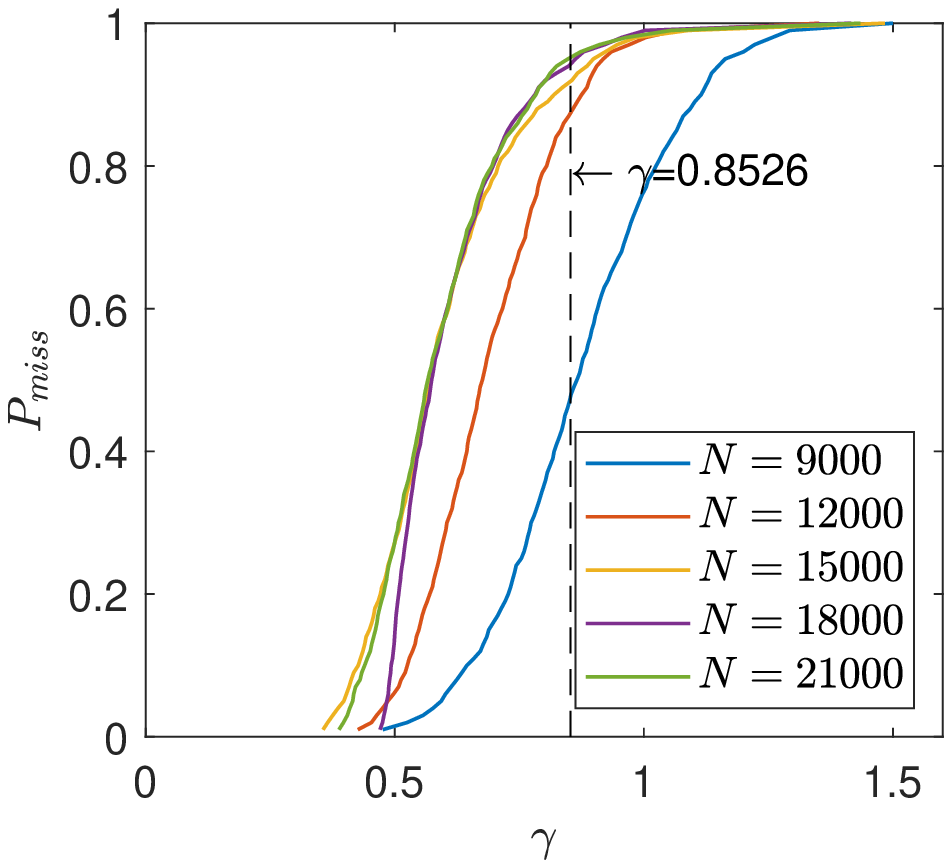}
\end{minipage}%
}%
\subfigure[]
{ 
\begin{minipage}[t]{0.5\linewidth}
\centering
\includegraphics[width=1.8in]{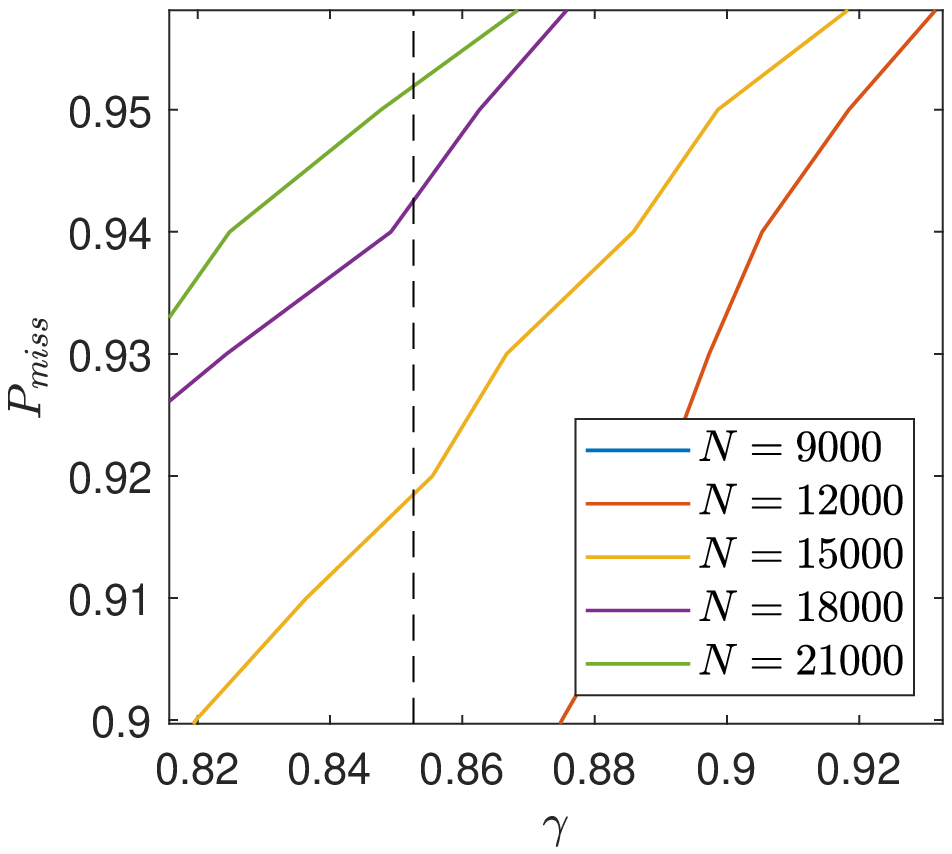}
\end{minipage}%
}
\centering
\caption{(a). Probability to pass the WLS state estimator  using different sample sizes; (b). Zoomed-in picture of (a). }
\label{different_sample_size}
\end{figure}

We focus on Attacking Region 1 when the target bus is bus 28. For comparison, we launch the same FDIA i.e, $\tilde V_{28}=0.8{V_{28}},\tilde \delta_{28}=\delta_{28}$ that is nevertheless designed using different sizes of PMU measurements.  Fig. \ref{different_sample_size} presents the probabilities to bypass the WLS state estimator when using $N= 21000$, $N= 18000$, $N= 15000$, $N= 12000$, $N= 9000$, i.e, $350$s, $300$s, $250$s, $200$s, $150$s, 60 Hz PMU data, respectively. Particularly, the probabilities to bypass the BDD are $95.2\%$, $94.6\%$, $91.9\%$, $87.5\%$, $47.5\%$, respectively.  The same threshold for the BDD as that in Section \ref{sectionexampleI} is applied, i.e., $95\%$-quantile $\gamma=0.8526$ from the residual distribution of the base case. \color{black} As expected, PMU measurements with a larger sample size yield more accurate line estimation and thus a higher success rate for the FDIA. 300s data seems to be a good trade-off between the sample size and the success rate. 

\subsection{Impacts of Different PMU Noises}
The Total Vector Error (TVE) is used to measure the PMU measurement noise as per  IEEE C37.118.1 Standard \cite{Power2011}. In this section, we test the performance of the proposed FDIA when using PMU measurements with different TVEs.  

\begin{figure} [htbp]
\centering
\subfigure[]
{ 
\begin{minipage}[t]{0.5\linewidth}
\centering
\includegraphics[width=1.8in]{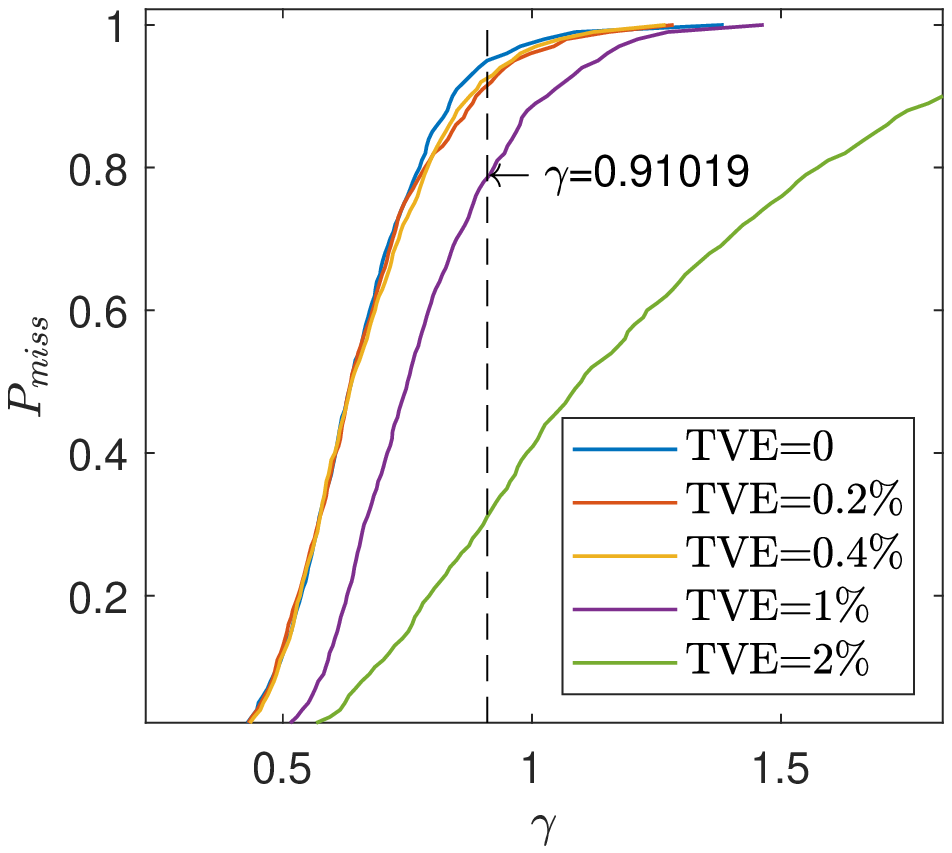}
\end{minipage}%
}%
\subfigure[]
{ 
\begin{minipage}[t]{0.5\linewidth}
\centering
\includegraphics[width=1.8in]{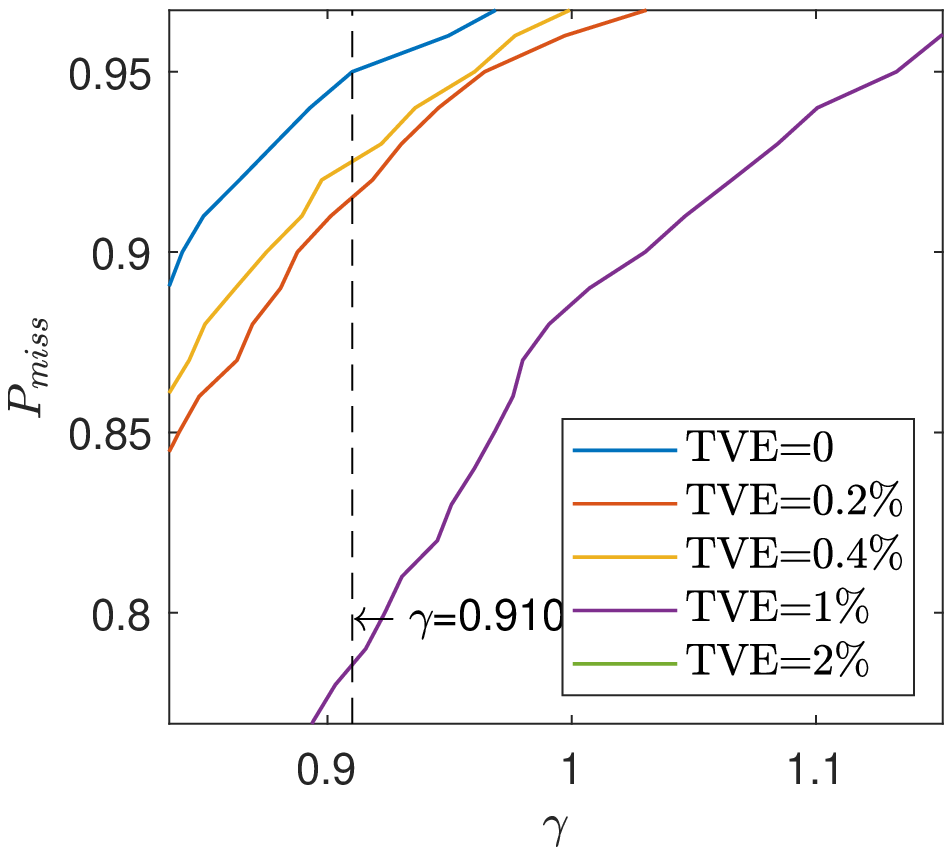}
\end{minipage}%
}%
\centering
\caption{(a). Probability to pass the WLS state estimator using different PMU noises; (b). Zoomed-in picture of (a). }
\label{attack under different PMU noise} 
\end{figure}

We focus on Attacking Region 1 when the target bus is bus 28. %Now, the time constants for the 
The time constants of the load inside the attacking region are $[{\tau _{{p_{26}}}},{\tau _{{p_{28}}}},{\tau _{{p_{29}}}}] =[3.24,0.8,5.14]$ s, $[{\tau _{{q_{26}}}},{\tau _{{q_{28}}}},{\tau _{{q_{29}}}}] = [6.814,1.2,0.894]$ s. The process noise intensities $\sigma_i^p$, $\sigma_i^q$ are set to 4. Assuming the intruder launches the voltage magnitude attack such that %we will test FDIAs under different PMU noises of 
${\tilde V_{28}} = 0.8{V_{28}}$, Fig. \ref{attack under different PMU noise} presents the probabilities to bypass the WLS state estimator for different TVE errors, which are obtained from 1000-sample Monte Carlo simulations. Particularly, the TVE errors are added to the measurements of phase angles.

The threshold for the BDD $\gamma= 0.9102$ is selected from the 95\%-quantile of the residual distribution of the base case, which is slightly different from that in the previous sections as the process noise intensities of loads in the base case are different. The probabilities to bypass the BDD for TVE=0.2\%, 0.4\%, 1\%, 2\% are 91.5\%, 92.5\%, 78.7\% and 31.2\%, respectively. As can be seen, the measurement noise may greatly affect the performance of the proposed FDIA, as the estimation of line parameters purely depends on PMU data. Filtering techniques and other advanced techniques to reduce the impacts of measurement noise will be investigated in the future.

\subsection{Impacts of Using Robust State Estimators}
\label{subsection:Robuststateestimation}

Although the traditional WLS estimator is widely used, robust state estimators such as weighted least absolute value (WLAV)\cite{Chen2015}, maximum exponential absolute value (MEAV)\cite{Chen2017}, maximum exponential square (MES)\cite{Wu2011} have recently been proposed to increase the accuracy and robustness of state estimation against bad data. In this section, we test the performance of the proposed FDIA against the MES robust state estimator proposed in\cite{Wu2011}.

\begin{figure} [htbp]

\centering
\subfigure[]
{ 
\begin{minipage}[t]{0.5\linewidth}
\centering
\includegraphics[width=1.8in]{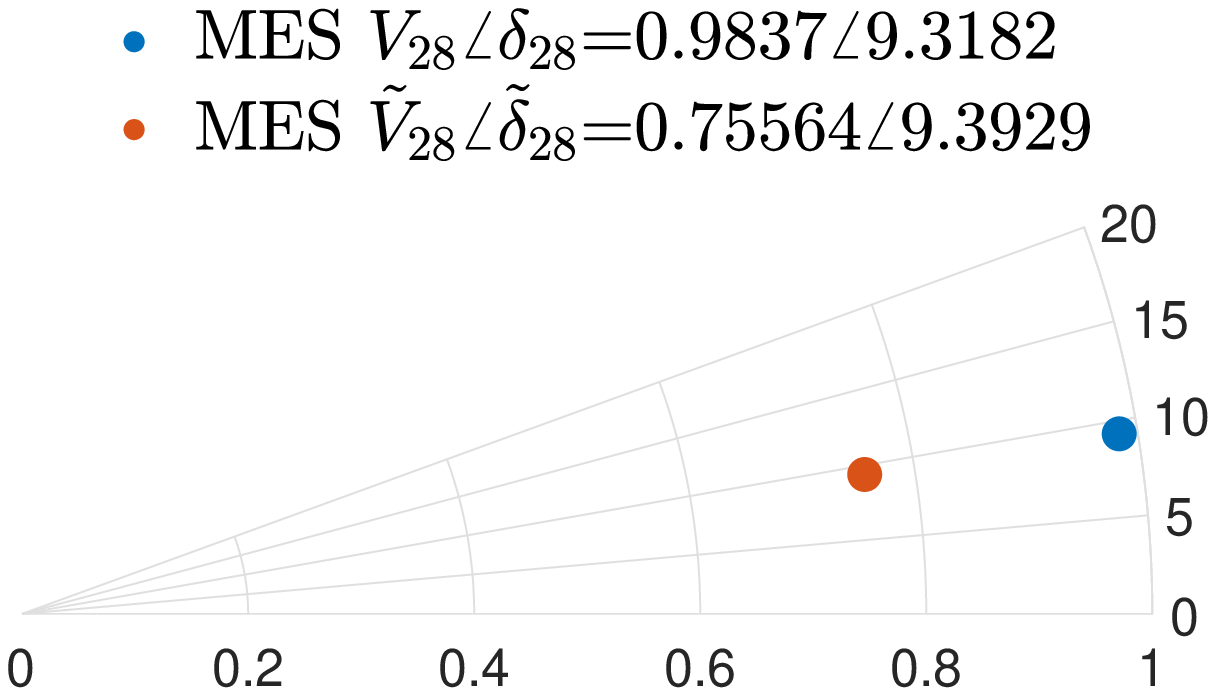}
\end{minipage}%
}%
\subfigure[]
{ 
\begin{minipage}[t]{0.5\linewidth}
\centering
\includegraphics[width=1.8in]{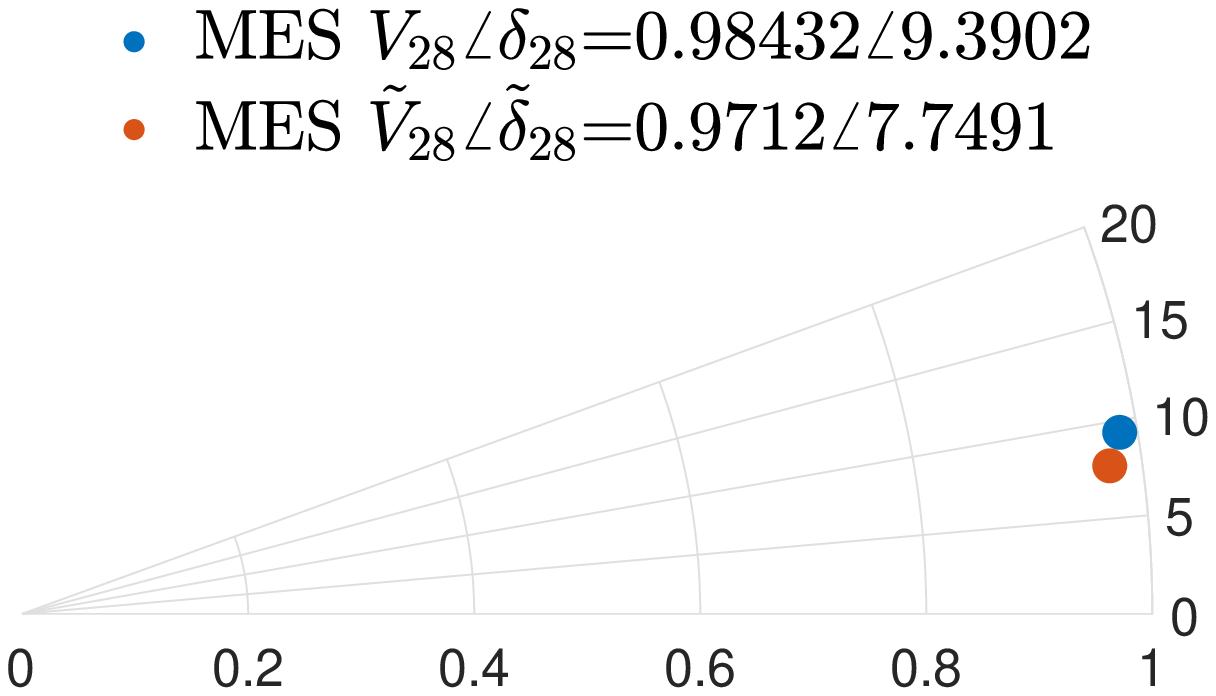}
\end{minipage}%
}
\centering
\caption{State estimation result of bus 28 under MES with FDIA ${{\rm{\tilde V}}_{28}} = 0.8{{\rm{V}}_{28}},{\tilde \delta _{28}} = {\delta _{28}}$ and ${{\rm{\tilde V}}_{28}} = 0.7{{\rm{V}}_{28}},{\tilde \delta _{28}} = {\delta _{28}}$} 
\label{MES_80_70} 
\end{figure}

Fig. \ref{MES_80_70} shows that the FDIA with distortion ${{\tilde V}_{28}} = 0.8{V_{28}},{{\tilde \delta }_{28}} = {\delta _{28}}$ can bypass the MES state estimation, %as well as the WLS state estimation (see Fig. \ref{residual_distribution_WLS}). However, if 
yet a larger size FDIA, e.g. ${{\tilde V}_{28}}= 0.7{V_{28}}, {{\tilde \delta }_{28}} = {\delta _{28}}$ cannot bypass the MES state estimation.  Note that both the attacks can bypass the WLS state estimator as presented in Fig. \ref{attack for V}. These results show that the implementation of MES state estimator may be effective in defending extremely large attack (30\%), yet it may still fail to defend against large deviation attacks ($20\%$).

\begin{figure} [htbp]
\centering
\subfigure[]
{ 
\begin{minipage}[t]{0.5\linewidth}
\centering
\includegraphics[width=1.8in]{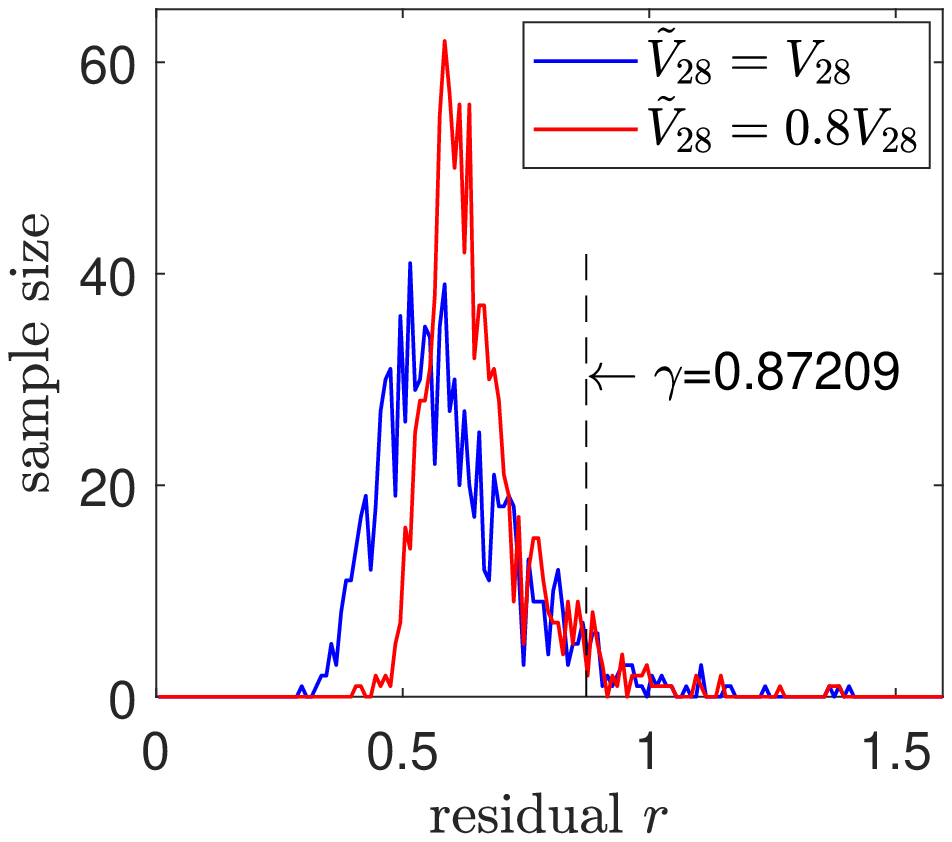}
\end{minipage}%
}%
\subfigure[]
{ 
\begin{minipage}[t]{0.5\linewidth}
\centering
\includegraphics[width=1.8in]{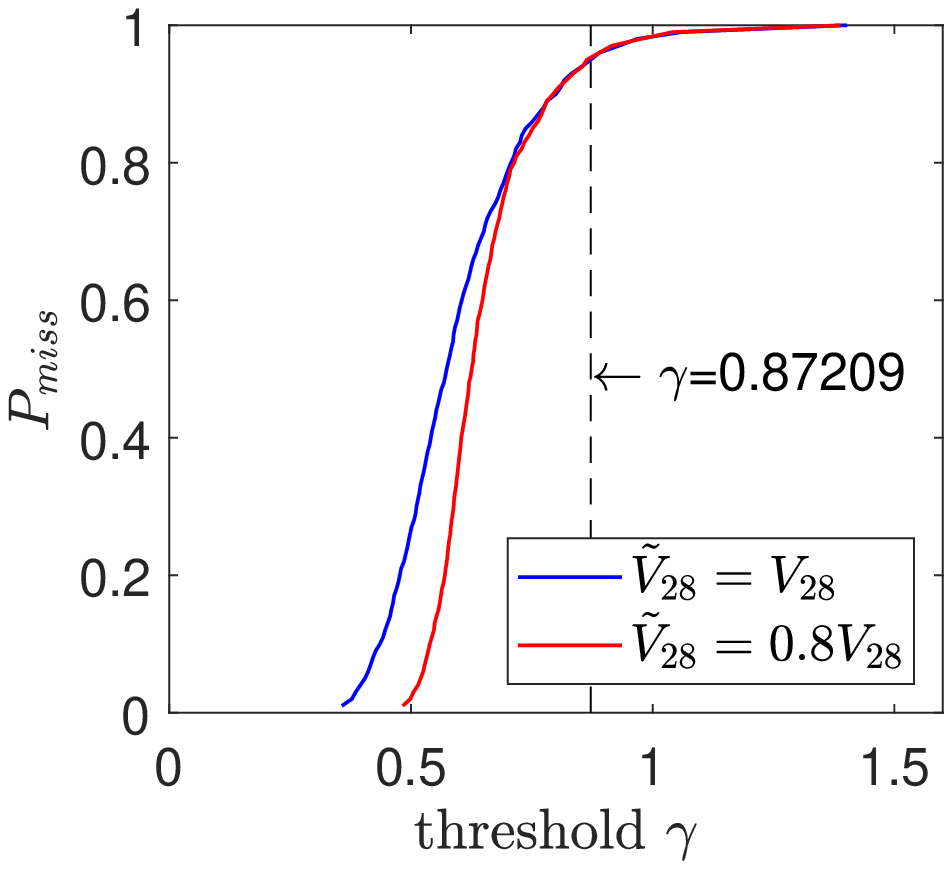}
\end{minipage}%
}
\centering
\caption{(a). The residual distribution before and after the attack $\tilde V_{28}=0.8{V_{28}}$ under MES; (b). The probability to bypass the MES estimator. } 
\label{residual_distribution_MES} 
\end{figure}

%To further demonstrate the feasibility of the proposed FDIA model under MES respectively, 
Furthermore, 1000 Monte Carlo simulations are run, in which the same attack $\tilde V_{28}=0.8{V_{28}}$ is implemented. 
%
%The MES state estimator is more sensitive to abnormal measurements, so it will slightly increase the residual distribution compared to WLS state estimator, thus increasing the threshold. 
%\color{black}
As shown in Fig. \ref{residual_distribution_MES}, although the probability curve of the residual of the MES state estimator shifts to the right after the FDIA, the probability of the proposed FDIA to bypass the MES state estimator is 95.2\% if 95\%-quantile $\gamma= 0.87209$ is selected to be the BDD threshold. Note that the threshold 95\%-quantile $\gamma= 0.87209$ is slightly different from the value in Section \ref{sectionexampleI}-Section \ref{section:sample size} since the residual distribution in the base case is different due to the implementation of the MES state estimation. \color{black} These results demonstrate that the proposed FDIA may still bypass the BDD with a high success rate even when the robust MES estimator is implemented at the control center. It is worth noting that the MEAV robust state estimator may outperform the MES state estimator in defending against the proposed FDIA as the normal measurements are weighted higher, which requires further study.

\subsection{Impacts of Integrating PMU measurements into the State Estimation} \label{subsection:PMUstateestimation}

\begin{figure} [htbp]
\centering
\subfigure[]
{ 
\begin{minipage}[t]{0.5\linewidth}
\centering
\includegraphics[width=1.8in]{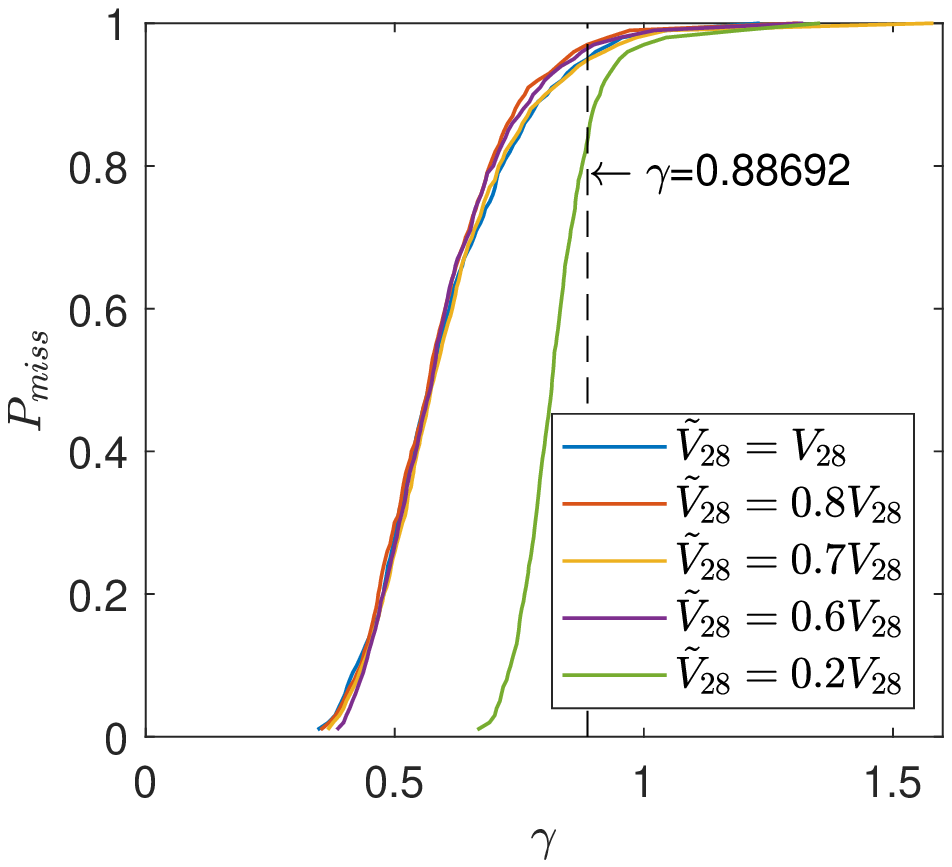}
\end{minipage}%
}%
\subfigure[]
{ 
\begin{minipage}[t]{0.5\linewidth}
\centering
\includegraphics[width=1.8in]{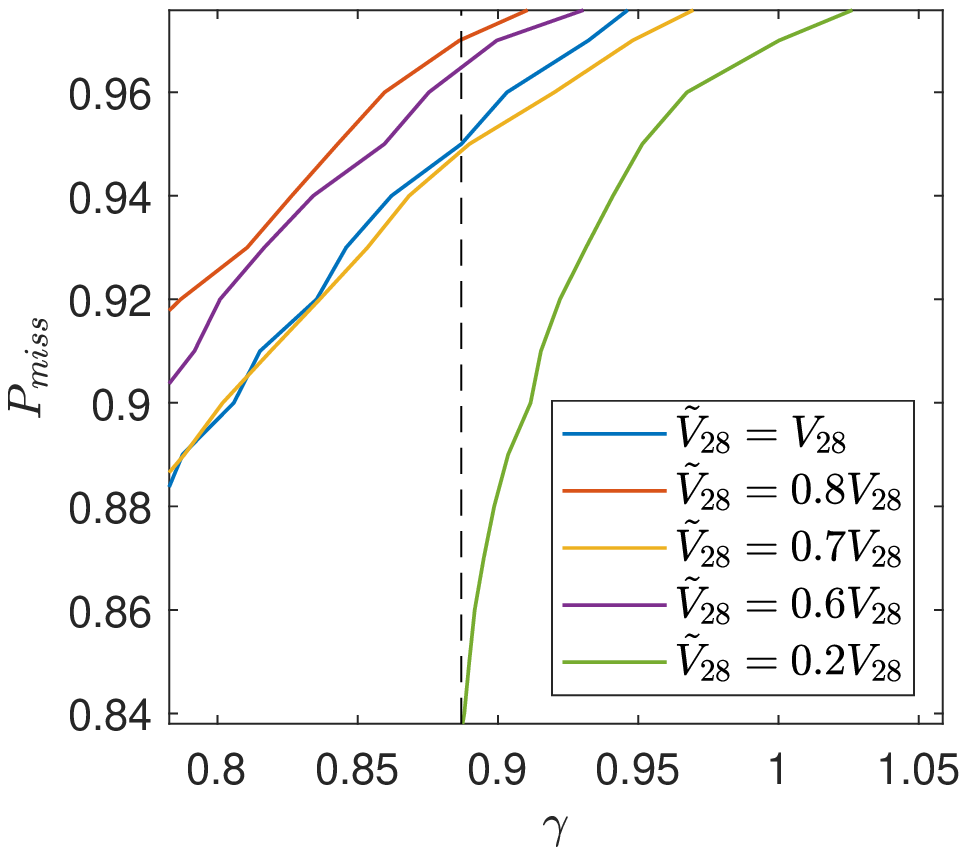}
\end{minipage}%
}
\centering
\caption{(a). Probability to pass the WLS state estimator integrated with both PMU voltage and current phasors; (b). Zoomed-in picture of (a). }
\label{state_estimation_with_PMU_voltage_current}
\end{figure}

We intend to investigate the impacts of integrating PMU data into the state estimation on the performance of the proposed FDIA model. Assuming there are 13 PMUs ($\frac{1}{3}$ of the number of total buses) installed at buses 3, 4, 7, 8, 12, 15, 16, 21, 24, 26, 28, 29, 39, respectively, the voltage phasors and current phasors of all PMUs are integrated into the state estimation. We focus on Attacking Region 1 when the target bus is bus 28. Before the attacks, ${V_{28}}\angle {\delta _{28}} = 0.98\angle {9.33^ \circ }$, then different attacks on the voltage magnitude of the target bus are tested through various Monte Carlo simulations using 1000 samples.

Fig. \ref{state_estimation_with_PMU_voltage_current} illustrates that the probabilities to bypass the BDD of ${\tilde V_{28}} = 0.8{V_{28}}$, ${\tilde V_{28}} = 0.7{V_{28}}$, ${\tilde V_{28}} = 0.6{V_{28}}$,  ${\tilde V_{28}} = 0.2{V_{28}}$ are 97.9\%, 94.9\%, 96.6\% and 83.8\%, respectively.  The threshold for the BDD is $\gamma=0.88692$ that is the 95\%-quantile of the residual distribution in the base case when PMU data is integrated into the WLS state estimator. \color{black} Compared to the results presented in Fig. \ref{attack for V}, after integrating PMU data into state estimation, the probabilities to bypass the BDD are comparable and even slightly higher except for the attack ${\tilde V_{28}} = 0.2{V_{28}}$. Although the detailed impacts (slight improvement or deterioration) are case-by-case, the bottom line is that integrating PMU measurements into the state estimation does not significantly affect the success rate of the proposed FDIA model. \color{black}

\section{Conclusions and Perspectives}\label{sectionconclusion}
In this paper, a novel targeted FDIA model against AC state estimation has been proposed, which, requiring no line parameters of a power network, can target specific states and launch large deviation attacks. Sufficient conditions for the proposed FDIA model to be perfect are also provided. Numerical studies for various attacks and different target buses show that the proposed FDIA model can always launch targeted large deviation attacks with a very high success rate. The proposed targeted FDIA model shows that {it may be even easier than we expected to} launch an FDIA since only limited PMU data is needed without any line parameters.
The proposed model and the results of the paper can be leveraged to design appropriate countermeasures in the near future.

\medskip
\normalem
\bibliographystyle{IEEEtran}
\bibliography{IEEEabrv,My_Collection.bib}

\vspace{-32pt} 

\begin{IEEEbiography}[{\includegraphics[width=1in,height=1.25in,clip,keepaspectratio]{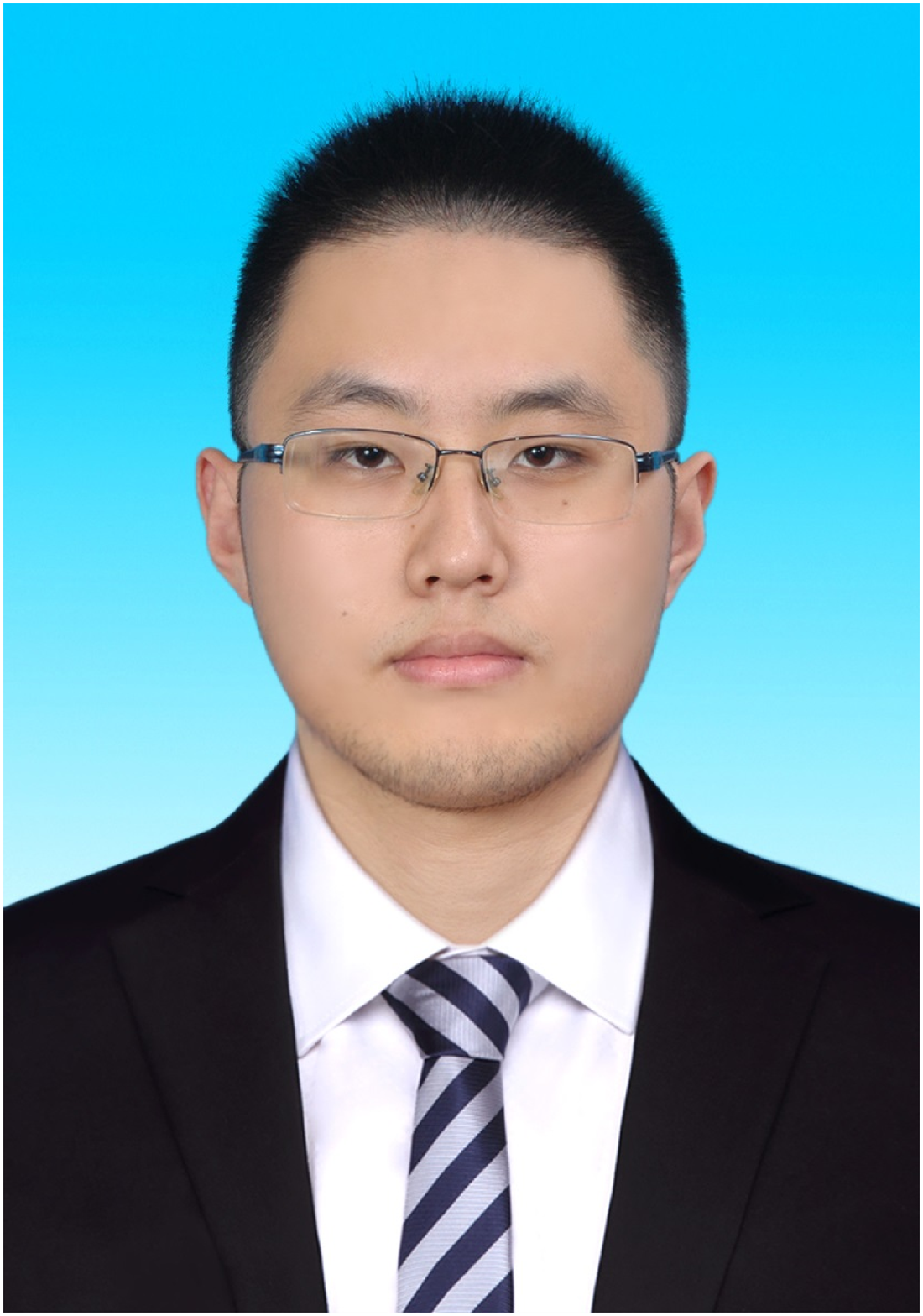}}]{Mingqiu Du} (Graduate Student Member, IEEE) received the B.Eng. and M.Sc. degrees in electrical engineering from Huazhong University of Science and Technology Wuhan, China, in 2015 and 2018, respectively. Since 2019, he has been pursuing the Ph.D. degree in Electrical Engineering with the Electric Energy Systems Laboratory at McGill University, Montreal, Canada. His research interests include power security and uncertainty quantification.
\end{IEEEbiography}

\begin{IEEEbiography}[{\includegraphics[width=1in,height=1.25in,clip,keepaspectratio]{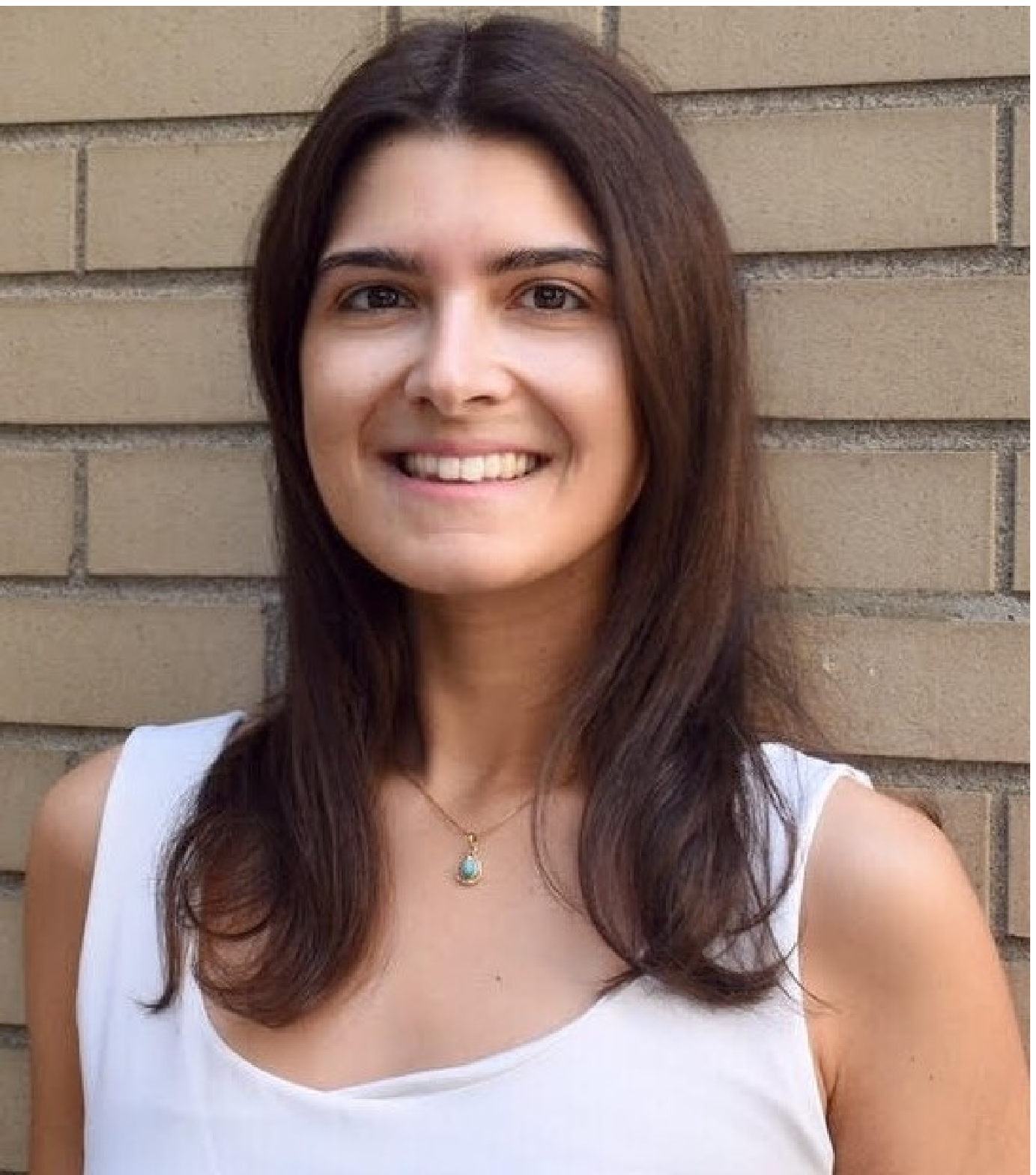}}]{Georgia Pierrou} (Student Member, IEEE) received the Diploma in Electrical and Computer Engineering from the National Technical University of Athens, Athens, Greece in 2017, and the Ph.D. degree in Electrical Engineering from McGill University, Montreal, Canada in 2021. Her research interests include power system dynamics, control and uncertainty quantification.
\end{IEEEbiography}

\begin{IEEEbiography}[{\includegraphics[width=1in,height=1.25in,clip,keepaspectratio]{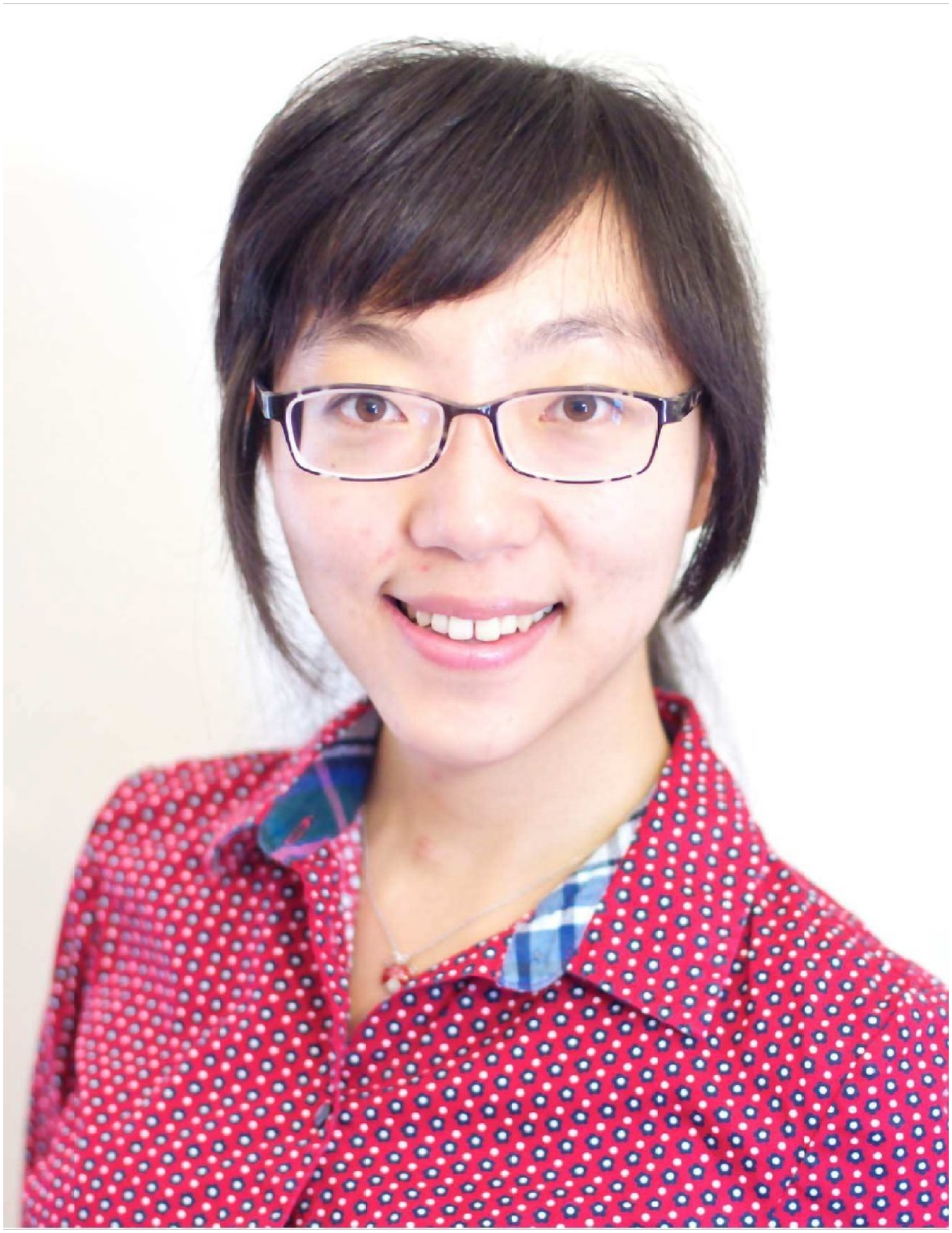}}]{Xiaozhe Wang} (Senior Member, IEEE) is currently an Assistant Professor in the Department of Electrical and Computer Engineering at McGill University, Montreal, QC, Canada. She received the Ph.D. degree in the School of Electrical and Computer Engineering from Cornell University, Ithaca, NY, USA, in 2015, and the B.S. degree in Information Science $\&$ Electronic Engineering from Zhejiang University, Zhejiang, China, in 2010. Her research interests are in the general areas of power system stability and control, uncertainty quantification in power system security and stability, and wide-area measurement system (WAMS)-based detection, estimation, and control. She is serving on the editorial boards of IEEE Transactions on Power Systems, Power Engineering Letters, and IET Generation, Transmission and Distribution.
\end{IEEEbiography}

\begin{IEEEbiography}[{\includegraphics[width=1in,height=1.25in,clip,keepaspectratio]{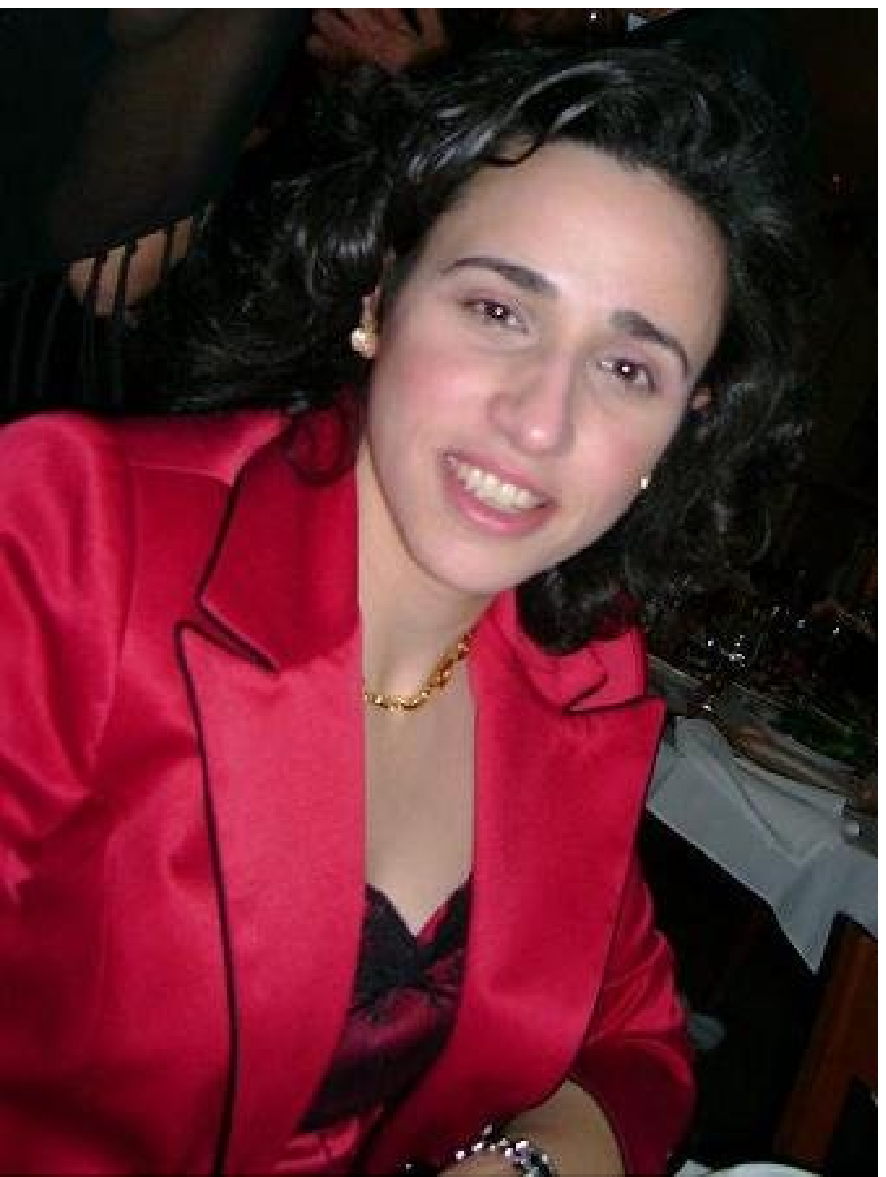}}]{Marthe Kassouf} received the B.Sc. and M.Sc. degrees in Computer Engineering from \'Ecole Sup\'erieure des Ing\'enieurs de Beyrouth, Lebanon (1997) and from \'Ecole Polytechnique de Montr\'eal, Canada (1999), respectively. In 2008, she received the Ph.D. degree in Electrical Engineering from McGill University, Canada. Since 2008, she has been working as a researcher at the Hydro Quebec Research Institute (IREQ), where she has been contributing to the implementation of different projects aiming at the enhancement of the information and telecommunications infrastructure supporting the power grid, mainly in the areas of wireless communication systems, time synchronization and cybersecurity. She has been the project manager for the cybersecurity research project at IREQ since 2018. She is also an adjunct professor at the Department of Electrical and Computer Engineering at McGill University. Active member in the Working Group 15 (WG15) of the International Electrotechnical Commission (IEC) Technical Committee (TC) 57 since 2015, she has been contributing to the development of IEC 62351 standards for the cybersecurity of power system information infrastructure. Her research interests are in telecommunication networks, time synchronization systems, power grid automation, and cybersecurity for smart grids.
\end{IEEEbiography}

\end{document}